\documentclass[prc,preprint,superscriptaddress,nofootinbib,showpacs]{revtex4}
\usepackage{amsfonts}
\usepackage{amssymb}
\usepackage{color,soul}
\usepackage{amsmath}
\usepackage{amsthm}
\usepackage{epsf}
\usepackage[dvips]{graphicx}

\begin{document}

\title{Extracting the longitudinal structure function $F_L(x,Q^2)$ at small $x$
from a Froissart-bounded
parametrization of $F_2(x,Q^2)$}

 \author{L.~P. Kaptari}
\affiliation{Institute of Modern Physics, Chinese Academy of Science, 509 Nanchang Road, 730000, Lanzhou, China}
\affiliation{Bogoliubov Laboratory of Theoretical Physics, Joint Institute for Nuclear Research, Dubna 141980, Russia}
\author{ A.V.~Kotikov}
\affiliation{Institute of Modern Physics, Chinese Academy of Science, 509 Nanchang Road, 730000, Lanzhou, China}
\affiliation{Bogoliubov Laboratory of Theoretical Physics, Joint Institute for Nuclear Research, Dubna 141980, Russia}
\author{N.Yu.~Chernikova}
\affiliation{Sunday school, 141980, Dubna, Russia }
\author {Pengming Zhang}
\affiliation{Institute of Modern Physics, Chinese Academy of Science, 509 Nanchang Road, 730000, Lanzhou, China}
\affiliation{ University of Chinese Academy of Sciences, Yuquanlu 19A, Beijing 100049, China}

\newcommand{\z}{&&\hspace*{-1cm}}
\newcommand{\zz}{&&\hspace*{-4cm}}
\newcommand{\ep}{\varepsilon}
\newcommand{\cita} [1] {$^{\hbox{\scriptsize \cite{#1}}}$}
\newcommand{\prepr}[1] {\begin{flushright} {\bf #1} \end{flushright} \vskip 1.5cm}
\newcommand{\bea}{\begin{eqnarray}}
\newcommand{\eea}{\end{eqnarray}}
\newcommand{\be}{\begin{equation}}
\newcommand{\ee}{\end{equation}}
\newcommand{\MSbar}{\overline{\rm MS}}
\newcommand{\as}{\alpha_s}
\newcommand{\asMZ}{\alpha_s(M^2_Z)}
\newcommand{\ar}{a_s}



\begin{abstract}
We present a method to extract, in the leading and next-to-leading order approximations, the
longitudinal deep-inelastic scattering structure function $F_L(x,Q^2)$ from the experimental data by
relying on a Froissart-bounded parametrization of the transversal structure function $F_2(x,Q^2)$
and, partially, on the  Dokshitzer-Gribov-Lipatov-Altarelli-Parisi equations. Particular attention is paid
on kinematics of low and ultra low values of the Bjorken variable $x$, $x\sim 10^{-5}\div 10^{-2}$.
Analytical expressions for  $F_L(x,Q^2)$ in terms of the effective parameters of the
 parametrization of $F_2(x,Q^2)$ are presented explicitly. We argue that the obtained structure functions $F_L(x,Q^2)$
within both, the leading and next-to-leading order approximations, manifestly obey the Froissart
boundary conditions. Numerical calculations and comparison with available data from ZEUS and H1-Collaborations
at HERA demonstrate that the suggested method provides reliable structure functions $F_L(x,Q^2)$ at low $x$
in a wide range of the momentum transfer $(1\, {\rm GeV^2} < Q^2\, < 3000\, {\rm GeV^2})$ and can be
applied as well in analyses of ultra-high energy processes with cosmic neutrinos.

\end{abstract}
\maketitle

\section{Introduction}
At small values of the Bjorken  variable $x$, the nonperturbative effects in
the deep inelastic structure functions (SF)  were expected to play a decisive role in describing
the corresponding cross sections.
 However,   it has been observed, cf. Ref.~\cite{CooperSarkar:1997jk},  that  even in the region of low momentum transfer $Q^2 \sim 1$ GeV$^2$,
 where traditionally the soft processes were considered to govern the cross sections,
 the perturbative QCD (pQCD) methods could  be still adequate in description of high energy processes, in particular, at relatively  low values of $x$: $10^{-5} \leq x \leq 10^{-2}$.
   This has been clearly  demonstrated in  early analyses of the {\it pre-}HERA data within approaches based on pQCD and on idea that the steep behavior of the momentum distributions at low $x$ can be generated purely dynamically, merely from measured valence densities, by utilizing QCD evolution equations~\cite{GluckZphys}.
 This idea has been confirmed by subsequent measurements of the structure function $F_2(x,Q^2)$ at $x> 10^{-2}$.
 At smaller  $x < 10^{-2}$, to achieve a better agreement with data, the dynamical parton distribution functions (PDF) require  additional
 fine tuning of the valence-like input parameters~\cite{Gluck:1998xa,Jimenez-Delgado:2014twa}.
 Such complementary tuning   results in rather stable parametrizations of PDF's in a broad  range of $Q^2$~\cite{Pisano:2008ms}. Furthermore, there are various groups (cf.~Refs. \cite{Jimenez-Delgado:2014twa,Dulat:2015mca}
 and references therin)  actively involved in extracting PDFs from experimental data, with particular attention paid   on description of low-x HERA and LHC data (cf. Refs.~\cite{Ball:2017otu,Abdolmaleki:2018jln,deOliveira:2017lop}). In most cases the extraction procedure
is supplemented  by  the small-$x$  Balitsky-Fadin-Kuraev-Lipatov (BFKL)
resummation~\cite{Fadin:1975cb}  of the  evolution equations  and DIS
 coefficient functions, thereby
leading to resummed PDF sets (for recent reviews see, for example, Refs.~\cite{Ball:2017otu,Abdolmaleki:2018jln}). It has been shown,  that the inclusion of BFKL  resummation
significantly improves the quantitative description of the small-$x$ and small-$Q^2$ HERA data, in
particular at next-to-leading order (NLO) and next-to-next-to-leading order (NNLO) approximation, both for the inclusive and for the charm
structure functions.
The resummation of logarithms at low $x$ also stabilizes the perturbative
expansion, and the resulting PDF's receive a specific shape, rising at small~$x$.


It should also  be noted that, at ultra low $x$, $x\to 0$, the pQCD evolution,
lead, nonetheless,  to a rather singular behaviour of PDFs
 (see e.g. Ref.~\cite{Kotikov:1998qt} and references therein quoted),  which is in a strong disagreement
 with the Froissart boundary conditions~\cite{Froissart:1961ux}.
 In Refs.~\cite{Berger:2007vf,Block:2011xb,Block:2014kza}   M.~M.~Block et al.   have suggested a new
 parametrization of the SF $F_2(x,Q^2)$
 which describes fairly well the available experimental data on the reduced cross sections and,
 at asymptotically low $x$, provides a behavior of  the hadron-hadron cross sections  $\sim \ln^2 s$
 at  large $s$, where $s$ is the Mandelstam variable denoting the square of the total
 invariant energy of the process,
 in a full accordance with the Froissart predictions~\cite{Froissart:1961ux}.
The most recent parametrization suggested in Ref.~\cite{Block:2014kza}
by  M.~M.~Block, L.~Durand and P.~Ha, in what follows  referred to as the BDH parametrization,
is also pertinent in investigations of  lepton-hadron processes at ultra-high energies ,
e.g.  the scattering of cosmic neutrinos from hadrons
~\cite{Block:2011xb,Block:2014kza,Illarionov:2011wc,Arguelles:2015wba,Bhattacharya:2016jce}.
Note that, in case of neutrino scattering other  SF's,   such as
the pure valence, $F_3(x,Q^2)$, and longitudinal,  $F_L(x,Q^2)$,
are relevant to describe the process. While at low values of $x$ the valence
structure function $F_3(x,Q^2)$ vanishes, the longitudinal   $F_L(x,Q^2)$
remains finite and can even be predominant in the cross section.
Thus, a theoretical analysis of the longitudinal SF $F_L(x,Q^2)$
at low $x$, in context of fulfilment of the  Froissart prescriptions,
is  of a great importance in treatments of ultra-high energy processes as well.

 Hitherto,  most theoretical analyses~\cite{Gandhi:1995tf,Arguelles:2015wba} of
 neutrino processes have been performed
in the leading order (LO) approximation, within which the  Callan-Gross relation is assumed
to be satisfied exactly,  i.e. the longitudinal structure function  $F_L=0$.
Beyond the LO the effects from  $F_L$ can be sizable, hence it can not be
longer neglected, cf. Ref.~\cite{Arguelles:2015wba,Anchordoqui:2019ufu}.

 In the present paper we present a  method of extraction of
the longitudinal  SF, $F_L(x,Q^2)$ in the kinematical region of low values of the  Bjorken variable $x$ from
the known structure function  $F_2^{\rm BDH}(x,Q^2)$ and known  derivative $dF_2^{\rm BDH}/d\ln(Q^2)$
 by  relying, with some extent, on the
   Dokshitzer-Gribov-Lipatov-Altarelli-Parisi (DGLAP)
 $Q^2$-evolution equations~\cite{Gribov:1972ri}. In our calculations we use the most recent
 version of the BDH-parametrization reported in Ref.~\cite{Block:2014kza}.
 In fact, the presented approach is a further development of the methods
 previously suggested in Refs.~\cite{Kotikov:1994vb,Kotikov:1994jh}
 to extract  some general characteristics of
  the gluon density and  longitudinal SF at low $x$ from the  the experimentally known
 SF $F_2(x,Q^2)$ and  logarithmic derivative $dF_2/d\ln Q^2$.
 The extraction procedure has been inspired by  the Altarelli-Martinelli formula \cite{Altarelli:1978tq}
  suggested to determine the gluon density from $F_L(x,Q^2)$, and improved in Ref.~\cite{Kotikov:1993xe}.

 In our case,  the SF  $F_2(x,Q^2)$
 is considered experimentally known being defined by the $F_2^{\rm BDH}(x,Q^2)$ parametrization~\cite{Block:2014kza},
 i.e. the $x$ and $Q^2$ dependencies of the transverse SF
 and the corresponding logarithmic derivative 
  are supposed to be known.
As a first step of the analysis, the method has been applied to extract
   $F_L(x,Q^2)$ in the LO. Results of such a procedure have been
briefly reported in Ref.~\cite{Kaptari:2018sxh}, where it
has been demonstrated that the extracted
structure function $F_L^{BDH}(x,Q^2)$ at moderate and low values of $x$ is in a reasonable good
 agreement with the available experimental data~\cite{Aaron:2009aa}.
 However, for  ultra low values of $x$ the agreement becomes less satisfactorily  and even rather poor
 in the limit $x\to 0$. This  serves as a clear indication that the LO analysis is not
 sufficient in the region $x\to 0$ and the NLO
 corrections become significant and are to
 be implemented in to the extraction procedure. Similar investigations of the longitudinal SF
 have been performed in Ref.~\cite{Boroun:2018lpz}.

In this  paper, we present in some details the LO analysis~\cite{Kaptari:2018sxh},
and provide further development of the method  extending it beyond the LO approximation
by considering and resumming the NLO corrections.

It is worth emphasizing that the NLO approximation for  $F_L(x,Q^2)$,
i.e. calculations up to $\alpha_s^2$-corrections,
corresponds to the NNLO
approximation for $F_2(x,Q^2)$  which,  in the LO, is $\propto \alpha_s^0$.
 Hence, in our approach it becomes
 possible  to perform NLO and NNLO analyses of the ultra-high energy ($\sqrt{s}\sim 1$~TeV) neutrino cross-sections similar to
existing NLO \cite{CooperSarkar:2007cv} and  NNLO ~\cite{Bertone:2018dse} investigations based on pQCD.
Such   analyses are rather important in view of
the recently appeared possibility of a direct comparison  with emerging data
from the IceCube Collaboration~\cite{Aartsen:2017kpd} (cf. also Ref. \cite{Bustamante:2017xuy}) and
anticipated data from the IceCube-Gen2~\cite{Aartsen:2014njl}, whose performance is much better and  which
 is awaited  to provide substantially more precise measurements  of
the neutrino-nucleon cross-section.

Our paper is organized as follows:\\
In Sec.~\ref{sec:2} we present the  basic formulae of  the approach.
The relevant system of equations to be used in extraction of the longitudinal
SF,  together with  the corresponding splitting functions
and coefficient functions are displayed  explicitly.   In Subsections~\ref{sec:mellin} and~\ref{anamol}
 we discuss the Mellin transforms of the transversal and longitudinal SF's in the LO and NLO approximations
 for momenta corresponding to low $x$.
Explicit expressions for the anomalous dimensions and Wilson coefficients in the LO for low $x$ are given as well.

In Sec.~\ref{sec:3} we write down  details of obtaining   all the needed, in the subsequent
calculations, quantities related to the BDH parametrization~\cite{Block:2014kza}, such as   the
corresponding  derivatives and  Mellin transforms within the considered kinematics and approximations.
Next two sections, Sec.~\ref{sec:LO} and Sec.~\ref{sec:NLO}, are entirely devoted
to  description of the gist of the mathematical
methods and manipulations used to calculate the Mellin transforms and their inverses
to find the longitudinal  SF in the LO, Sec.~\ref{sec:LO},  and NLO, Sec.\ref{sec:NLO},  approximations.
Numerical results  for  the extracted $F_L(x,Q^2)$ in the LO and NLO, together with comparisons with the experimental data
from the  H1-Collaboration, are presented in Sec.~\ref{sec:results}, where we discuss the $Q^2$ and $x$-dependencies of the extracted SF $F_L(x,Q^2)$ and the  ratio $R_L(x,Q^2)$  of the longitudinal to transversal cross sections within the LO and NLO  approximations.
Conclusions and summary are summarized on Sec.~\ref{sec:con}. Eventually, the most cumbersome expressions are relegated to Appendices~\ref{App:B}~and~\ref{app:mel}.

\section{Basic formulae}\label{sec:2}

In view of  at low values of $x$  the non-singlet quark distributions become negligibly small in comparison
with  the singlet distributions, in the present analysis they are  disregarded.
 Then, the  transverse $F_2(x,Q^2)$ and longitudinal  $F_L(x,Q^2)$ structure functions  are expressed
 solely via
the  singlet   quark   and gluon  densities $xf_a(x,Q^2)$ (hereafter $a=s,g$ and $k=2,L$) as
\be
F_k(x,Q^2)  = e \sum_{a=s,g} \biggl[ B_{k,a} (x) \otimes xf_a(x,Q^2)
 \biggr], \label{Fk}
\ee
  where $e$ is the average charge squared, $
e=\frac{1}{f} \sum^f_{i=1} e_i^2 \equiv \frac{e_{2f}}{f}
$
with  $f$ as the number of considered  flavors,
$q^2=-Q^2$ and $x=Q^2/2pq$ ($p$ being the momentum of the  nucleon)
denote  the  momentum transfer and the Bjorken scaling variable, respectively.
The quantities  $B_{k,a} (x)$  are the known Wilson coefficient functions.
 In Eq.~(\ref{Fk}) and throughout the rest of the paper, the symbol $\otimes$ is used for a
 shorthand notation of the convolution formula, i.e.
 $f_1(x)\otimes f_2(x)\equiv \int\limits_x^1\displaystyle\frac{dy}{y}f_1(y) f_2\left(\displaystyle\frac{x}{y}\right)$.

According to the DGLAP   $Q^2$-evolution equations~\cite{Gribov:1972ri} the leading
twist quark, $xf_s(x,Q^2)$,  and gluon, $xf_g(x,Q^2)$, distributions obey the following system of
integro-differential  equations
 \be
\frac{d (xf_a(x,Q^2))}{dlnQ^2}  = - \frac{1}{2}
\sum_{a,b=s,g} P^{(0)}_{ab}
(x) \otimes  xf_b(x,Q^2),
\label{DGLAPgen}
\ee
where $P_{ab} (x)$ $(a,b=s,g)$  are the corresponding splitting functions.

Within the pQCD, and up to the NLO corrections, the coefficient functions $B_{k,a} (x)$ and the splitting functions $P_{ab} (x)$
read as
\bea
&&B_{2,s}(x)=  \delta(1-x) + a_s(Q^2) \, B^{(1)}_{2,s}(x), \label{B2} \\
&&B_{2,g}(x)=  a_s(Q^2) \, B^{(1)}_{2,g}(x), \label{B2a}\\
&&B_{L,a}(x)= a_s(Q^2) \, B^{(0)}_{L,a}(x) + a_s^2(Q^2) \, B^{(1)}_{L,a}(x), \label{BL} \\
&&P_{a,b}(x)= a_s(Q^2) \, P^{(0)}_{a,b}(x) + a_s^2(Q^2) \, P^{(1)}_{a,b}(x) \, ,\label{Pab}
\eea
 where $a_s(Q^2)=\alpha_s(Q^2)/4\pi$ is the QCD  running coupling, which, for convenience, includes in to
 its definition an additional
factor of $4\pi$ in comparison with the standard notation. In the above equations and hereafter
 the superscripts $(0,1)$ mark
the corresponding order of the perturbation theory: $(0)$ for LO and  $(1)$ for NLO.

Inserting Eqs.~(\ref{B2})-(\ref{Pab}) in to Eqs.~(\ref{Fk})-(\ref{DGLAPgen})
 the final NLO system of equations for the sought PDF's becomes
\bea
\frac{d (xf_g(x,Q^2))}{dlnQ^2}  &=& -\frac{a_s(Q^2)}{2}  \biggl[
\Bigl( P^{(0)}_{gg} (x) +  a_s(Q^2)\, \tilde{P}^{(1)}_{gg} (x) \Bigr)
\otimes xf_g(x,Q^2) +\nonumber \\
&&
 e^{-1} \Bigl( P^{(0)}_{gs} (x) + a_s(Q^2)\,  \tilde{P}^{(1)}_{gs} (x) \Bigr)
\otimes F_2(x,Q^2)+~ O(a^3_s) \biggr]
\label{fg} \\
\frac{dF_2(x,Q^2)}{dlnQ^2}  &=& -\frac{a_s(Q^2)}{2} \biggl[e
 \Bigl( P^{(0)}_{sg} (x) + a_s(Q^2)\, \tilde{P}^{(1)}_{sg} (x) \Bigr)
\otimes xf_g(x,Q^2)  \nonumber \\
&& +
\Bigl( P^{(0)}_{ss} (x) +  a_s(Q^2)\, \tilde{P}^{(1)}_{ss} (x) \Bigr)
\otimes F_2(x,Q^2) +~ O(a^3_s) \biggr] , \label{dF2} \\
F_L(x,Q^2)  &=& a_s(Q^2) \biggl[ e
\Bigl( B^{(0)}_{L,g} (x) + \tilde{B}^{(1)}_{L,g} (x) \Bigr) \otimes xf_g(x,Q^2)
+\Bigl( B^{(0)}_{L,q} (x) \nonumber \\
&& a_s(Q^2)\,  \tilde{B}^{(1)}_{L,q} (x) \Bigr)
\otimes F_2(x,Q^2) +~ O(a^3_s) \biggr] , \label{FL.1}
\eea
where, for  brevity, the following notations have been employed
\bea
&&\tilde{P}^{(1)}_{sg} (x) = P^{(1)}_{sg} (x) + B^{(1)}_{2,s} (x) \otimes P^{(0)}_{sg} (x)
+ B^{(0)}_{2,g} (x) \otimes \Bigl( 2\beta_0 \delta(1-x) + P^{(0)}_{gg} (x) -  P^{(0)}_{ss} (x) \Bigr), \nonumber \\
&&\tilde{P}^{(1)}_{ss} (x) = P^{(1)}_{ss} (x) + 2\beta_0 B^{(1)}_{2,s} (x) \otimes \delta(1-x)
+ B^{(1)}_{2,g} (x) \otimes P^{(0)}_{gq} (x),  \nonumber \\
&&\tilde{P}^{(1)}_{gs} (x) = P^{(1)}_{gs} (x) - B^{(1)}_{2,s} (x) \otimes P^{(0)}_{gs} (x),~~
\tilde{P}^{(1)}_{gg} (x) = P^{(1)}_{gg} (x) - B^{(1)}_{2,g} (x) \otimes P^{(0)}_{gs} (x) \label{tPgq} \\
&&\tilde{B}^{(1)}_{L,g} (x) = B^{(1)}_{L,g} (x) - B^{(1)}_{2,g} (x) \otimes B^{(0)}_{L,g} (x),~~
\tilde{B}^{(1)}_{L,s} (x) = B^{(1)}_{L,s} (x) - B^{(1)}_{2,s} (x) \otimes B^{(0)}_{L,s} (x) \label{tPLq}
\eea
with $\beta_0$ and $\beta_1$ as the first two coefficients of the QCD $\beta$-function
\be
\beta_0= \frac{1}{3} \Bigl( 11C_A -2f \Bigr),~~ \beta_1= \frac{1}{3} \Bigl( 34C_A^2 -2f (5C_A + 3C_F)  \Bigr)\, .
\label{beta0}\
\ee
In Eq.~(\ref{beta0}) $C_F=(N_c^2-1)/(2N_c)$  and $C_A=N_c$  are the Casimir operators in the fundamental and adjoint
representations of the $SU(N_c)$ color group, respectively.
Within  QCD $N_c=3$, hence  $C_F=4/3$ and $C_A=3$.

Few remarks are in order here.
As known~\cite{Gribov:1984tu,Zhu:1998hg},  equation (\ref{fg}) in its actual form leads to a  too
singular behaviour of the gluon distribution at small x, violating the Froissart boundary restrictions. One can go beyond the perturbative theory and try to cure the problem by adding in the r.h.s. of Eq.~(\ref{fg})
terms proportional to $(xf_g)^2$ which make the distribution (hence, the corresponding cross-sections~\cite{Fiore:2004nt}) less singular at the origin
and can, in principle, reconcile it  with the Froissart requirements.
A  detailed inspection of Eq.~(\ref{fg}) in context of implementation of
additional modifications to fulfill  the Froissart conditions is beyond
the scope of the present paper and in what follows we omit it in our analysis. However, the gluon
distribution  originating from the omitted  Eq.~(\ref{fg}) and entering in to the
remaining equations  (\ref{dF2}) and (\ref{FL.1}) is supposed to have the
correct asymptotic behavior, i.e. to be of the  same LO form as the
 BDH parametrization of the $F_2^{\rm BDH}(x,Q^2)$.
This conjectures has been confirmed in previous analysis~\cite{Chernikova:2016xwx}
where the early parametrization of $F_2(x,Q^2)$~\cite{Block:2011xb} has been
employed to determine the gluon density within the LO.
Then, Eq.~(\ref{dF2}) with the known $F_2^{\rm BDH}(x,Q^2)$, can be considered as the
definition of the gluon density $xf_g^{\rm BDH}(x,Q^2)$
in the whole kinematical interval, cf. Ref.~\cite{Chernikova:2016xwx}. Consequently,
in the system (\ref{dF2})-(\ref{FL.1}) of two equations with two unknown distributions one can eliminate the
 gluon part and   solve the remaining equation  with respect to the longitudinal $F_L(x,Q^2) $
and express it via the known paramterization of $F_2(x,Q^2) $.
With these statements,  now we are in a position to solve Eqs.~(\ref{dF2}) and (\ref{FL.1}) and  to extract
the desired longitudinal SF.
Notice that, albeit  at the first glance  the above equations are
relatively simple, direct solving of (\ref{dF2})-(\ref{FL.1}) actually turns out to be a rather complicate
and cumbersome procedure. One can substantially simplify the calculations by
considering Eqs.~(\ref{dF2})-(\ref{FL.1})
 in the space of Mellin momenta, and taking advantage of the fact the
convolution form $f_1(x)\otimes f_2(x)$ in  $x$ space
becomes merely a product of individual Mellin transforms of the corresponding
functions in the  space of Mellin momenta. Consequently, all our further calculations
we perform  in  Mellin space.

\subsection{Mellin transforms} \label{sec:mellin}
The Mellin transform of the PDF's entering in to  Eqs.~(\ref{dF2})-(\ref{FL.1}) are defined as

\bea
&&M_k(n,Q^2) = \int^1_0 dx x^{n-2} F_k(x,Q^2),~~~
M_a(n,Q^2) = \int^1_0 dx x^{n-1} f_a(x,Q^2)\, ,
\label{Mg}\\
&&\gamma^{(i)}_{ab} (n) = \int^1_0 dx x^{n-2}
P^{(i)}_{ab} (x),~~~ B^{(i)}_{k,a} (n) = \int^1_0 dx x^{n-2}
B^{(i)}_{k,a} (x),
\label{gamma}
\end{eqnarray}
where, as before, $ a,b= s,g $ and $k= 2,L $.
Then, after some algebra, the Mellin transforms of
Eqs.~(\ref{dF2})-(\ref{FL.1}) read as
\bea
\frac{dM_2(n,Q^2)}{dlnQ^2}  &=& -\frac{a_s(Q^2)}{2} \biggl[
e \Bigl( \gamma^{(0)}_{sg} (n) + a_s(Q^2) \tilde{\gamma}^{(1)}_{sg} (n) \Bigr)
M_g(n,Q^2) + \Bigl( \gamma^{(0)}_{ss} (n) \nonumber \\
&&+ a_s(Q^2) \tilde{\gamma}^{(1)}_{ss} (n) \Bigr)
M_2(x,Q^2)
+~ O(a^3_s)
 \biggr] , \label{dM2}  \\
M_L(n,Q^2)  &=& a_s(Q^2) \biggl[e
\Bigl( B^{(0)}_{L,g} (n) + a_s(Q^2) \tilde{B}^{(1)}_{L,g} (n) \Bigr) M_g(x,Q^2)\nonumber \\
&&+
\Bigl( B^{(0)}_{L,s} (n) + a_s(Q^2) \tilde{B}^{(1)}_{L,s} (n) \Bigr) M_2(x,Q^2) +~ O(a^3_s)
 \biggr] , \label{FL.2}
\eea
where the anomalous dimensions $\gamma^{(i)}_{ab} (n)$ and the Wilson  coefficients $B^{(i)}_{L,a} (n)$ ($i=0,1$) are
\bea
&&\tilde{\gamma}^{(1)}_{sg} (n) = \gamma^{(1)}_{sg} (n) + B^{(1)}_{2,s} (n) \gamma^{(0)}_{sg} (n)
+ B^{(0)}_{2,g} (n) \Bigl( 2\beta_0 + \gamma^{(0)}_{gg} (n) -  \gamma^{(0)}_{ss} (n) \Bigr), \nonumber \\
&&\tilde{\gamma}^{(1)}_{ss} (n) = \gamma^{(1)}_{ss} (x) + 2\beta_0 B^{(1)}_{2,s} (n)
+ B^{(1)}_{2,g} (n) \gamma^{(0)}_{gq} (n),  \nonumber \\
&&\tilde{\gamma}^{(1)}_{gs} (n) = P^{(1)}_{gs} (n) - B^{(1)}_{2,s} (n) \gamma^{(0)}_{gs} (x),~~
\tilde{\gamma}^{(1)}_{gg} (n) = \gamma^{(1)}_{gg} (n) - B^{(1)}_{2,g} (n) \gamma^{(0)}_{gs} (n) \label{tPgqn} \\
&&\tilde{B}^{(1)}_{L,g} (n) = B^{(1)}_{L,g} (n) - B^{(1)}_{2,g} (n) B^{(0)}_{L,g} (n),~~
\tilde{B}^{(1)}_{L,s} (n) = B^{(1)}_{L,s} (n) - B^{(1)}_{2,s} (n) B^{(0)}_{L,s} (n) \label{tPLqn}
\eea

The explicit expressions for the
anomalous dimensions $\gamma^{(i)}_{ab} (n)$ ($i=0,1$) and the LO
 Wilson  coefficients $B^{(0)}_{L,a} (n)$   can be found  in Ref.~\cite{Floratos:1981hs}, whereas
the NLO  part, $B^{(1)}_{L,NS} (n)$ and $B^{(1)}_{L,a} (n)$,
has been reported, for the first time,  in Ref.~\cite{Kazakov:1987jk}. Unfortunately, there are several
  misprints in the above mentioned references. Erratum for
 Ref.~\cite{Kazakov:1987jk} can be found  in, e.g. Refs.~\cite{Kazakov:1992xj,Zijlstra:1992qd}.
 Yet, in Ref.~\cite{Floratos:1981hs} a  factor two in the LO coefficients $B^{(0)}_{L,a} (n)$ is missed.
Observe that  as mentioned above, the Mellin transform significantly
simplifies our  calculations  for, the complicated integro-differential system of
equations~(\ref{dF2})-(\ref{FL.1}) in $x$-space, is translated in to
 a relatively simple system (\ref{dM2})-(\ref{FL.2}) of pure algebraic equations.
 Now, we solve the system (\ref{dM2})-(\ref{FL.2}) w.r.t  the longitudinal Mellin momentum
 $M_L(n,Q^2)$ and  express it  through known momentum   $M_2(x,Q^2)$ and the derivative $dM_2(n,Q^2)/dlnQ^2$
\begin{eqnarray}
&&M_L(n,Q^2)  = - 2 \frac{B^{(0)}_{L,g} (n) + a_s(Q^2) \tilde{B}^{(1)}_{L,g} (n)}{\gamma^{(0)}_{sg} (n) + a_s(Q^2)
\tilde{\gamma}^{(1)}_{sg} (n)} \, \frac{dM_2(n,Q^2)}{dlnQ^2} +
\Biggl[\Bigl( B^{(0)}_{L,s} (n) + a_s(Q^2) \tilde{B}^{(1)}_{L,s} (n) \Bigr)
\nonumber \\
&&
- \Bigl( B^{(0)}_{L,g} (n) + a_s(Q^2) \tilde{B}^{(1)}_{L,g} (n) \Bigr) \,
\frac{\gamma^{(0)}_{ss} (n)  + a_s(Q^2) \tilde{\gamma}^{(1)}_{ss} (n)}{\gamma^{(0)}_{sg} (n)
+ a_s(Q^2) \tilde{\gamma}^{(1)}_{sg} (n)} \Biggr]\, M_2(x,Q^2) +~ O(a^3_s)    .
 \label{FL.2a}
\end{eqnarray}

\subsection{Anomalous dimensions and coefficient functions} \label{anamol}
Here below  we present the explicit expressions for the LO ingredients only. The corresponding
 expressions for the NLO corrections are rather cumbersome and, as already mentioned, can be found in
 Refs.~\cite{Kazakov:1987jk,Kazakov:1992xj,Zijlstra:1992qd,Floratos:1981hs}. Explicitly, the LO
 anomalous dimensions $\gamma^{(0)}_{ab} (n)$   and the Wilson  coefficients $B^{(0)}_{L,a} (n)$,   are
 as follows:
 \bea
\z \gamma^{(0)}_{sg} (n) = - \frac{4f (n^2+n+2)}{n(n+1)(n+2)},~~~
\gamma^{(0)}_{ss} (n) ~=~ 8C_F \, \biggl[S_1(n)
-\frac{3}{4} - \frac{1}{2n(n+1)}
\biggr], \label{ana} \\
\z \gamma^{(0)}_{ga} (n) = - \frac{4C_F (n^2+n+2)}{(n-1)n(n+1)},~~~
\gamma^{(0)}_{ss} (n) ~=~ 8C_A \, \biggl[S_1(n)
-\frac{1}{(n-1)n} - \frac{1}{(n+1)(n+2)}\biggr] + 2\beta_0,
\nonumber \\
\z B^{(0)}_{L,g} (n) =  \frac{8f}{(n+1)(n+2)},~~~
B^{(0)}_{L,q} (n) ~=~  \frac{4C_F}{(n+1)}.
\label{BL1}
\eea
In Eqs.~(\ref{ana})-(\ref{BL1}) we introduce,  and shall widely use throughout the rest
of the paper,   the notion of the so-called nested sums, defined as
\be
S_{\pm i}(n) = \sum_{m=1}^{n} \frac{(\pm 1)^m}{m^i},~~
S_{\pm i, j}(n) = \sum_{m=1}^{n} \frac{(\pm 1)^m}{m^i} \, S_{j}(m).
\label{Sij}
\ee

Note that  in previous calculations~\cite{Floratos:1981hs,Kazakov:1987jk,Kazakov:1992xj} the notion
of nested sums (\ref{Sij}) has not been yet incorporated. Instead, other notations related
to the nested sums of negative indices have been used
\bea
&&S'_m\left(\frac{n}{2}\right) = 2^{m-1} \Bigl( S_{m}(n) + S_{-m}(n)\Bigr),~~ \tilde{S}_m(n)=S_{-2,1}(n)
\mbox{ (Ref.~\cite{Floratos:1981hs})} \, ,\nonumber \\
&&K_m(n) = -S_{-m}(n),~~ Q(n)=-S_{-2,1}(n) \mbox{ (Refs. \cite{Kazakov:1987jk,Kazakov:1992xj})} \, .
\label{others}
\eea

Coming back to the Mellin transforms (\ref{dM2}), (\ref{FL.2}) and (\ref{FL.2a}),
we recall that we are interested in
investigation of the PDF's in the region of low  $x$. In the Mellin space it corresponds to small momenta,
and at extremely low $x$, it suffices  to restrict the analysis with the first momentum
and to study the solutions  of (\ref{dM2}), (\ref{FL.2}) and (\ref{FL.2a}) for $n=1+\omega$ at $\omega\to 0$.
  The nested sums $S_m(n)$ (here $n$ is not mandatorily an integer) at positive indices $m$
 can be expressed via the familiar  Riemann $\Psi$-functions $\Psi(n)$
as
\be
S_1(n)=\Psi(n+1) - \Psi(1),~~ S_m(n)=\zeta_m - \sum_{l=0}^{\infty} \frac{1}{(n+l+1)^m} ,\quad (m > 1),
\label{Sm}
\ee
where   $\zeta_m $ are Euler constants  and
the last series in the r.h.s. of (\ref{Sm})  is  related with the $m$-th  derivative of the $\Psi$-function.

A more complicated situation occurs for negative indices $m<0$
for which the analytical continuation of the nested sums depends on the parity of the starting value of $n$.  Since the anomalous
dimensions (\ref{ana}) have been calculated for even $n$, in what follows we employ the analytical
continuation of $S_{-m}(n)$ and $S_{-m,k}(n)$    starting from even values of $n$.
The result is~\cite{Kotikov:2005gr}
\bea
 \z S_1(n)=-\ln 2 - \sum_{l=0}^{\infty} \frac{(-1)^{l+1}}{n+l+1},~
S_{-m}(n)=\zeta_{-m} - \sum_{l=0}^{\infty} \frac{(-1)^{l+1}}{(n+l+1)^m} \, (m \geq 2),~ \zeta_{-m}= \left(1-2^{-m}\right) \zeta_{m}, \nonumber \\ \z
S_{-m,k}(n)=\sum_{l=0}^{\infty} \frac{(-1)^{l+1}}{(l+1)^m} S_k(l+1)
- \sum_{l=0}^{\infty} \frac{(-1)^{l+1}}{(n+l+1)^m} S_k( n+l+1)
 \, ,
\label{-Sm}
\eea
where the above series are well defined for any $n$ including  noninteger  values.

In what follows we are interested in quantities which contribute
to NLO at low $x$, i.e. in the
initial series $S_{-2}(n)$,  $S_{-3}(n)$ and $S_{-2,1}(n)$,
 anomalous dimensions and the coefficient functions, at $n=1+\omega$. In the vicinity of
 $\omega=0$  we have

\be
S_{-2}(1+\omega)= 1-\zeta_2 + O(\omega),~~ S_{-3}(1+\omega)= 1-\frac{3}{2} \zeta_3 + O(\omega),~~
S_{-2,1}(1+\omega)= 1-\frac{5}{4} \zeta_3 + O(\omega)\, .
\label{AnaCo}
\ee

\section{BDH-like results} \label{sec:3}

We reiterate that, our analysis is based on Eqs.~(\ref{dF2})-(\ref{FL.1}) or, equivalently, on
   Eqs.~(\ref{dM2}), (\ref{FL.2}) and (\ref{FL.2a}), where the structure function $F_2(x,Q^2)$ is supposed to be known and determined
 from the existing experimental data. In the present paper we employ the   BDH
 parametrization~\cite{Block:2014kza},
 obtained from a combined fit of the H1 and ZEUS collaborations data~\cite{Aaron:2009aa} in a   range of
the  kinematical variables $x$ and $Q^2$, $x < 0.01$ and  $0.15$ GeV$^2 < Q^2 < 3000$ GeV$^2$.
The explicit expression for the BDH parametrization reads as
\bea
F_{2}^{\rm BDH}(x,Q^2) = D(Q^2)
(1-x)^{\nu} \, \sum_{m=0}^2 A_m(Q^2) L^m
\label{parametrizationBDH}
\eea
where the dependence on the effective parameters is encoded in  $D(Q^2)$ and $A_m(Q^2)$
\bea
D(Q^2) = \frac{Q^2(Q^2 + \lambda M^2)}{(Q^2 + M^2)^2},\quad
 A_0(Q^2)=a_{00}+ a_{01}L_2,~~ A_i(Q^2) = \sum_{k=0}^2 a_{ik} \, L_2^k,~~
i=(1,2),~~
\label{n9.0}
\eea
with the logarithmic terms $L$ as
\bea
&&L= \ln \frac{1}{x} + L_1,~~ L_1= \ln \left(\frac{Q^2}{Q^2 + \mu^2}\right),~~
L_2=\ln\left(\frac{Q^2+\mu^2}{\mu^2}\right).
\label{dependencies}
\eea

The performed fit of the experimental data~\cite{Aaron:2009aa} provided the following values of the effective
parameters~\cite{Block:2014kza}:
\be
\mu^2=2.82 \pm 0.29 {\mbox GeV}^2,~~ M^2=0.753 \pm 0.008 {\mbox GeV}^2,~~ \nu=11.49 \pm 0.99,~~
\lambda = 2.430 \pm 0.153,
\label{n9.1}
\ee
and
\bea
&&a_{00} = 0.255 \pm 0.016,~~
a_{01} \cdot 10^{1} = 1.475 \pm 0.3025,~~ \nonumber \\
&&a_{10} \cdot 10^{4} = 8.205 \pm 4.620,~~
a_{11} \cdot 10^{2} = -5.148 \pm 0.819,~~
a_{12} \cdot 10^{3} = -4.725 \pm 1.010, \nonumber \\
&&a_{20} \cdot 10^{3} = 2.217 \pm 0.142,~~
a_{21} \cdot 10^{2}=
1.244 \pm 0.0860,~~
a_{22} \cdot 10^{4} =
5.958 \pm 2.320 \, .
\label{n10}
\eea
Notice that, the BDH parametrization (\ref{parametrizationBDH}) is written in the $\left( x-Q^2\right)$ space, whereas  the equation (\ref{FL.2a})
is in  space of Mellin momenta. Hence, before proceeding
with  consideration of  Eq.~ (\ref{FL.2a}),
we   transform the BDH parametrization in to Mellin space.
Then, in the  transformed parametrization   $M_2(n,Q^2)$      we consider  the first momenta
$n=1+\omega$ and   take the   limit $\omega \to 0$, which  corresponds to low $x$,  and obtain
\be
M_{2}^{\rm BDH}(n,Q^2) = D(Q^2) \, \sum_{m=0}^2 A_m(Q^2) P_m(\omega,\nu,L_1) + O(\omega) \equiv  D(Q^2) \,
\hat{M}_{2}^{\rm BDH}(n,Q^2) \, , \label{M2BDH}
\ee
where the quantity $P_m(\omega,\nu,L_1)$ stands for the approximate expression of the integral
\be \int^{1}_0 dx x^{\omega-1} (1-x)^{\nu} L^k(x) = P_k(\omega,\nu,L_1) + O(\omega).
\label{IntMain}
\ee
The explicit expression for the integral~(\ref{IntMain}) is relegated in   Appendix~\ref{App:B}. It should be noted that
$P_k(\omega,\nu,L_1)$, besides the finite part at $\omega\to 0$, contains also   negative powers of
$\omega$, i.e. is a singular  function at $\omega= 0$. Namely these singularities are of our interests in further
procedure of solving  Eqs.~(\ref{dF2})-(\ref{FL.2a}). The strategy is as follow: we disregard the finite part of
$M_{2}^{\rm BDH}(n,Q^2)$ keeping only the singular terms. Then the same procedure we repeat for the
longitudinal momentum $M_L^{\rm BDH}(n,Q^2)$, Eq. (\ref{FL.2a}), and equate the coefficients in front of
each singularity. In such a way we obtain the representation for the longitudinal SF.
Actually, in our calculations, we analyse the singularities in a slightly different
way by  using  some specific properties of the expansions over $\omega$, avoiding
direct comparison of the singular coefficients, see below.
At $\omega \to 0$ the integral  (\ref{IntMain}) becomes independent up on $\nu$  and the singular part
of  $\hat{M}_{2}^{\rm BDH}(n,Q^2)$ can written as

\be
\hat{M}_{2}^{\rm BDH}(n,Q^2) =   \sum_{m=0}^2 A_{m}(Q^2) P_m^{\rm sing}(\omega,L_1) + O(\omega^0),
 \label{M2BDHsing}
\ee
where
\be
P_0^{\rm sing}(\omega)=\frac{1}{\omega} \, ,~~
P_1^{\rm sing}(\omega,L_1)= \frac{1}{\omega^2} + \frac{L_1}{\omega} \, ,~~
P_2^{\rm sing}(\omega,L_1)= \frac{2}{\omega^3} + \frac{2L_1}{\omega^2} + \frac{L_1^2}{\omega} \, .
\label{PiSing}
\ee

Note that  $P_i^{\rm sing}(\omega)$ ($i=1,2,3$) in the limit $\omega \to 0$ satisfy the
following  useful recurrent  relations
\bea
&&\omega \, P_0^{\rm sing}(\omega)= O(\omega^0),\quad
\omega \, P_1^{\rm sing}(\omega)= P_0^{\rm sing}(\omega) + O(\omega^0),~~ \nonumber \\
&&\omega \, P_2^{\rm sing}(\omega)= 2 P_1^{\rm sing}(\omega) + O(\omega^0),~~ \omega^2 \, P_2^{\rm sing}(\omega)=
2 P_0^{\rm sing}(\omega) + O(\omega^0),
\label{PiSing.1}
\eea
which are widely used subsequently.

\subsection{Derivation of $M_{2}^{\rm BDH}(n,Q^2)$}
To conclude the low-x analysis one needs the explicit expressions for
  $dM_2^{\rm BDH}(n,Q^2)/d\ln Q^2$ in Eq. (\ref{FL.2a}), i.e.
  the derivatives of the corresponding ingredients
\bea
&&\frac{d}{d \ln Q^2} \, M_{2}^{\rm BDH}(n,Q^2) = \frac{d D(Q^2)}{d \ln Q^2} \,  \sum_{m=0}^2 A_{m}(Q^2) P_m^{\rm sing}(\omega,L_1) \nonumber \\
&&+ D(Q^2) \, \sum_{m=0}^2 \frac{d A_{m}(Q^2)}{d \ln Q^2} \, P_m^{\rm sing}(\omega,L_1)
+ D(Q^2) \, \sum_{m=0}^2 A_{m}(Q^2) \frac{d P_m^{\rm sing}(\omega,L_1)}{d \ln Q^2} + O(\omega^0)
\, .
\label{DifM2}
\eea

Noticing that
\be
\frac{d}{d \ln Q^2} \, L_2 = \frac{Q^2}{Q^2+\mu^2},~~
\frac{d}{d \ln Q^2} \, L_1 = \frac{\mu^2}{Q^2+\mu^2},~~
\frac{d}{d \ln Q^2} \, D =  \frac{M^2Q^2((2-\lambda)Q^2 + \lambda M^2)}{(Q^2 + M^2)^3} \equiv \tilde{D},
\label{Dif}
\ee
we can write
\bea
\z \frac{d}{d \ln Q^2} \, A_{m} = \frac{Q^2}{Q^2+\mu^2} \, \overline{A}_{m},~~
\overline{A}_{m}= a_{m1} + 2 a_{m2} L_2,~~a_{02}=0,~~
\label{DifA} \\[2mm]
&& \frac{d}{d \ln Q^2} \, P_m^{\rm sing}(\omega,L_1) = \frac{Q^2}{Q^2+\mu^2} \, \overline{P}_m^{\rm sing}(\omega,L_1),~~
\overline{P}_m^{\rm sing}(\omega,L_1)= (m-1) P_{m-1}^{\rm sing}(\omega,L_1) \, .
\label{DifP}
\eea

Collecting all results together, we have
\be
\frac{d}{d \ln Q^2} \, M_{2}^{\rm BDH}(n,Q^2) = \sum_{m=0}^2 \hat{A}_{m}(Q^2) P_m^{\rm sing}(\omega,L_1) + O(\omega^0)
\, , \label{dM2BDH}
\ee
where
\be
 \hat{A}_2 =  \tilde{A}_2,~~ \hat{A}_1 =  \tilde{A}_1 + 2D A_2 \, \frac{\mu^2}{Q^2+\mu^2},~~
 \hat{A}_0 =  \tilde{A}_0 + D A_1 \, \frac{\mu^2}{Q^2+\mu^2} \,
\label{hatAi}
\ee
and
\be
\tilde{A}_i = \tilde{D} A_{i} + D \overline{A}_i \, \frac{Q^2}{Q^2+\mu^2} \, .
\label{tAi}
\ee

\section{LO analysis}
\label{sec:LO}
Herein bellow we present in details the extraction of the
 longitudinal SF, $F_L^{\rm BDH}(x,Q^2)$,  in the LO approximation.
  In the LO  Eq.~(\ref{FL.2a}) reads as
\be
M_{L,\rm LO}(n,Q^2)  =  -2 \frac{B^{(0)}_{L,g} (n)}{\gamma^{(0)}_{sg} (n)} \frac{dM_2(n,Q^2)}{dlnQ^2}
+  a_s(Q^2)
\tilde{B}^{(0)}_{L,s} (n) M_2(x,Q^2)   , \label{FL.LO.1}
\ee
where
\be
\tilde{B}^{(0)}_{L,s} (n)= B^{(0)}_{L,s} (n) - B^{(0)}_{L,g} (n) \frac{\gamma^{(0)}_{ss} (n)}{ \gamma^{(0)}_{sg} (n)} \, . \label{tBOLs}
\ee

The next step is to consider Eq.~(\ref{FL.LO.1}) at $n=1+\omega$ and
to perform the series expansion   of the anomalous dimensions and  the coefficient functions
about  $\omega=0$. Restricting the expansion up to terms $\propto\omega^2$ we obtain
 \bea &&
 \frac{B^{(0)}_{L,g}(1+\omega)}{e \gamma^{(0)}_{sg}(1+\omega)}  =
- \frac{1}{2} \, \left(1+\frac{\omega}{4}  - \frac{7}{16} \omega^2 \right); 
\quad
 B^{(0)}_{L,s}(1+\omega) = 2 C_F \, \left(1-\frac{\omega}{2}  +
 \frac{\omega^2}{4}  \right), 
 \nonumber\\
&& \gamma^{(0)}_{ss}(1+\omega) = 8C_F \omega  \, \left( \zeta(2)-\frac{5}{8} +  \left(\frac{9}{16} - \zeta(3) \right)\omega \right) \, .
 \label{AnExp}
\eea
Observe that, in the above equation (\ref{FL.LO.1}) both, the momentum
 $M_2(x,Q^2)$ and the derivative $dM_2(n,Q^2)/dlnQ^2$
are of a similar form, cf. Eqs.~(\ref{M2BDHsing}) and (\ref{dM2BDH}), so that the r.h.s. of
 (\ref{FL.LO.1})  with the expansions (\ref{AnExp})
 can be represented as
 $$\sum\limits_{m=0}^2 A_{m} P_m^{\rm sing}\times \,\,
 \left(\phantom{\frac11\!\!\!\!}{\rm a\,\, cubic\,\, polynomial\,\, in }\,\ \omega\right).$$ This substantially simplifies the calculations  since, in such a case,  we can apply
  the recurrent relations
 (\ref{PiSing.1}) to find the desired coefficients. For instance, for any (known) series of the form
 \be
 T(\omega)= T_0 + T_1 \omega + T_2 \omega^2 + O(\omega^3)
 \label{T}
\ee
 one easily obtains the result
 \bea
 T(\omega) \sum_{m=0}^2 A_{m} P_m^{\rm sing} &=& T_0 \sum_{m=0}^2 A_{m} P_m^{\rm sing} + T_1
 \Bigr(2A_2P_2^{\rm sing} + A_1P_1^{\rm sing} \Bigr) + 2 T_2 A_2P_1^{\rm sing} + O(\omega^0) \nonumber \\
 &=&
 \sum_{m=0}^2 \overline{\overline{A}}_{m} P_m^{\rm sing} + O(\omega^0) , \label{M2.T}
\eea
where
\be
\overline{\overline{A}}_2 = T_0 A_2,~~ \overline{\overline{A}}_1 = T_0 A_1 + 2T_1 A_2, ~~ \overline{\overline{A}}_0 = T_0 A_0 + T_1 A_1 + 2T_2 A_2 .
\label{M2.T1}
\ee
It is seen that it is sufficient to determine the coefficients $T_0$, $T_1$ and $T_2$
 to find the solution of the system,  avoiding in such a way,  lengthy and cumbersome calculations.
 In our case the explicit expressions
for $T_i$  can be inferred directly from Eq. (\ref{AnExp}).
Then the coefficients
$C_m$ of the LO  expansions of the longitudinal Mellin  momenta
\be
 M_{L, \rm LO}^{\rm BDH}(n,Q^2) = \sum_{m=0}^2 C_m P_m^{\rm sing} + O(\omega^0), \label{MLFu}
\ee
explicitly read  as
\bea
C_{2} &=& \hat{A}_2 +  \frac{8}{3} a_s D A_2 \, ,~~
C_{1} = \hat{A}_1 +  \frac{1}{2} \, \hat{A}_2 +
  \frac{8}{3} a_s D \Bigl(A_1 + \Bigl(4\zeta_2 -\frac{7}{2}\Bigr) A_2 \Bigr) ,
  \nonumber \\
C_{0} &=& \hat{A}_0 +  \frac{1}{4} \, \hat{A}_2 -\frac{7}{8} \, \hat{A}_2
+  \frac{8}{3} a_s D \left(A_0 + \Bigl(2\zeta_2 -\frac{7}{4}\Bigr) A_1 +
 \left(\zeta_2-4\zeta_3 +\frac{17}{8}\right) A_2 \right) \,  .
\label{C0}
\eea
  It is worth mentioning   once more that,  the above results for   $C_i$ ($i=0,1,2$)
have been obtained in a straight way avoiding   direct comparisons of singularities in
(\ref{FL.LO.1}), as previously reported  in~\cite{Kaptari:2018sxh,Chernikova:2016xwx}.
Another important result is that  the  adopted   BDH  parametrization
for the $F_2(x,Q^2)$  led directly to   the same BDH-like form of
the   longitudinal structure functions   $F_{L, \rm LO}^{\rm BDH}(x,Q^2)$ which, consequently, also
 obey the Froissart boundary condition
\be
F_{L, \rm LO}^{\rm BDH}(x,Q^2) = (1-x)^{\nu_{L}} \, \sum_{m=0}^2 C_m(Q^2) L^m,
\label{n9.01}
\ee
where $\nu_{L}=\nu$ and, according to   the quark counting rules \cite{Matveev:1973ra}. 

\section{NLO analysis
}
\label{sec:NLO}
 In this section we discuss   the   longitudinal structure function within
 the NLO approximation. As before, particular attention is paid to   the region of
 asymptotically small  $x\to 0$ in context with the  Froissart boundary.

Multiplying  both sides of Eq.~(\ref{FL.2a}) by the
factor $\Bigl(1 + a_s(Q^2) \Bigl[\delta^{(1)}_{sg} (n) - R^{(1)}_{L,g} (n)\Bigr]\Bigr)$, we have
\bea
&&\Bigl(1 + a_s(Q^2) \Bigl[\delta^{(1)}_{sg} (n) - R^{(1)}_{L,g} (n)\Bigr] \Bigr) M_L(n,Q^2)=
-2 \frac{B^{(0)}_{L,g} (n)}{\gamma^{(0)}_{sg} (n)} \frac{dM_2(n,Q^2)}{dlnQ^2} \nonumber \\
&&+ a_s(Q^2)\left(\tilde{B}^{(0)}_{L,s} (n) + a_s(Q^2) \tilde{B}^{(1)}_{L,s} (n) \right)
M_2(n,Q^2)+~ O(a^3_s),
 \label{FL.4n}
\eea
where
\be
\delta^{(1)}_{sa}(n) = \frac{\tilde{\gamma}^{(1)}_{sa}(n)}{\gamma^{(0)}_{sg} (n)},~~
R^{(1)}_{L,a} (n) = \frac{\tilde{B}^{(1)}_{L,a} (n)}{B^{(0)}_{L,a} (n)}
\label{tdelta}
\ee
and
\be
 \tilde{B}^{(1)}_{L,s} (n) = B^{(0)}_{L,s} (n) \Bigl(R^{(1)}_{L,s} (n) + \delta^{(1)}_{sg}(n) - R^{(1)}_{L,g} (n)\Bigr) -
 B^{(0)}_{L,g} (n) \delta^{(1)}_{ss}(n) \, .
 \label{tRLs}
 \ee

Using equations (\ref{FL.LO.1}) and (\ref{FL.4n}) together, it is straightforward to obtain
\bea \!\!\!\!\!\!
\Bigl(1 + a_s(Q^2) \Bigl[\delta^{(1)}_{sg} (n) - R^{(1)}_{L,g} (n)\Bigr]
\Bigr) M_L(n,Q^2)  =
M_{L,\rm LO}(n,Q^2)+ 
a^2_s(Q^2)\tilde{B}^{(1)}_{L,s} (n)M_2(n,Q^2) \, .
\label{FL.6n}
\eea

 When computing the inverse Mellin transform of (\ref{FL.6n}) we
 employ the fact that in the l.h.s.  of (\ref{FL.6n}), up to $ O(a^3_s)$ corrections,  we can write
\be
a_s(Q^2) \Bigl[\delta^{(1)}_{sg} (n) - R^{(1)}_{L,g} (n)\Bigr] M_L(n,Q^2) = a_s(Q^2) \Bigl[\delta^{(1)}_{sg} (n) - R^{(1)}_{L,g} (n)\Bigr]
M_{L,\rm LO}(n,Q^2) +~ O(a^3_s)
\, ,
\label{fgNLO.2}
\ee
where the LO  momentum  $M_{L,\rm LO}(n,Q^2)$ has already been   calculated
in the previous section, cf. Eq.~(\ref{MLFu}).

Prior to proceed  with the inverse Mellin transforms,
 it is convenient to extract the singular structure of the NLO coefficients
$\delta^{(1)}_{sg} (n)$, $R^{(1)}_{L,g} (n)$ and $\tilde{B}^{(1)}_{L,s} (n)$. We have
\be
\delta^{(1)}_{sg}(n)
= \frac{\hat{\delta}^{(1)}_{sa}}{\omega}+ \overline{\delta}^{(1)}_{sa}(1+\omega) ,~~
R^{(1)}_{L,g}(n) = \frac{\hat{R}^{(1)}_{L,g}}{\omega}+ \overline{R}^{(1)}_{L,g}(1+\omega) ,~~
\tilde{B}^{(1)}_{L,s}(n) = \frac{\hat{B}^{(1)}_{L,g}}{\omega}+ \overline{B}^{(1)}_{L,g}(1+\omega) \, ,
\label{tdeltaEx}
\ee
with ($\omega\to 0$)
\bea
&& \hat{B}^{(1)}_{L,s}=\frac{20}{3} C_F \Bigl(3C_A - 2f\Bigr),~~\overline{B}^{(1)}_{L,s}(1)= 8C_F \left[ \frac{25}{9} f-
 \frac{449}{72} C_F + (2C_F-C_A) \left(\zeta_3+2\zeta_2 -\frac{59}{72} \right)\right], \nonumber \\
&&\hat{\delta}^{(1)}_{sg}=\frac{26}{3} C_A,\quad\overline{\delta}^{(1)}_{sg}(1)=3C_F-\frac{347}{18} C_A,\quad
\hat{R}^{(1)}_{L,g}= -\frac{4}{3} C_A,\quad\overline{R}^{(1)}_{L,g}(1)=-5C_F-\frac{4}{9} C_A \, .
\label{tdelta.1}
\eea

Then, the equation (\ref{FL.6n}) can be rewritten in the following form
\bea \!\!\!\!\!\!
&&M_L(n,Q^2) + \frac{a_s(Q^2)}{\omega} \Bigl[\hat{\delta}^{(1)}_{sg} (n) - \hat{R}^{(1)}_{L,g} (n)\Bigr]
 M_{L,\rm LO}(n,Q^2)  =
\Bigl(1 - a_s(Q^2) \Bigl[\overline{\delta}^{(1)}_{sg} (n) - \overline{R}^{(1)}_{L,g} (n)\Bigr]
\Bigr) + \nonumber \\&&
\times M_{L,\rm LO}(n,Q^2) +
a^2_s(Q^2) \left[\frac{\hat{B}^{(1)}_{L,s}}{\omega}
+
\overline{B}^{(1)}_{L,s} (n) \right] M_2(n,Q^2) +~ O(a^3_s) \, .
\label{FL.6na}
\eea

Now the  inverse Mellin transforms of the
last equations can be easily performed (see also Appendix~\ref{app:mel}). The result is
\bea &&\hspace*{-10mm}
{F}^{\rm BDH}_L(x,Q^2) + \frac{a_s(Q^2)}{3} L_C \left[{\hat{\delta}^{(1)}_{sg}-\hat{R}^{(1)}_{L,g}} \right]
 {F}^{\rm BDH}_{L,\rm LO}(x,Q^2)=\nonumber \\ && \hspace*{-10mm}
\left[1- a_s(Q^2)\left(\overline{\delta}^{(1)}_{sg}(1) - \overline{R}^{(1)}_{L,g}(1) \right)\right]
 {F}^{\rm BDH}_{L,\rm LO}(x,Q^2)
- a_s^2(Q^2)\left[\frac{\hat{B}^{(1)}_{L,s}}{3} L_A
+ \overline{B}^{(1)}_{L,s}(1) \right] M^{\rm BDH}_2(x,Q^2),
 \label{FL.6nb}
\eea
where
\be
L_A=L+\frac{A_1}{2A_2}; \quad L_C=L+\frac{C_1}{2C_2} \, .
\label{LB}
\ee

With the considered accuracy the obtained equation (\ref{FL.6nb}) can be rewritten as
\bea
&&\left[1 + \frac13 a_s(Q^2) L_C \left({\hat{\delta}^{(1)}_{sg}-\hat{R}^{(1)}_{L,g}} \right)
\right]
 {F}^{\rm BDH}_L(x,Q^2)=
\left[ 1-a_s(Q^2)\left( \overline\delta^{(1)}_{sg}-\overline R_{L,g}\right)\right]{F}^{\rm BDH}_{L,\rm LO}(x,Q^2)
\nonumber \\&&
- a_s^2(Q^2)\left[ \frac13 \hat{B}^{(1)}_{L,s} L_A
+ \overline{B}^{(1)}_{L,s}(1) \right]
M^{\rm BDH}_2(x,Q^2) +~ O(a^3_s) \, . \label{FL.7n}
\eea

Eventually, the final expression for the longitudinal SF  $\hat{F}^{\rm BDH}_L(x,Q^2)$
reads as
\bea
{F}^{\rm BDH}_L(x,Q^2) &=& \frac{1}{\left[1 + \frac13 a_s(Q^2) L_C \left(\hat{\delta}^{(1)}_{sg}(1)-\hat{R}^{(1)}_{L,g}\right)
 \right]}
\left\{\phantom{\!\!\!\!\!\! \frac{1^1}{1}}
 \left[ 1-a_s(Q^2)\left( \overline\delta^{(1)}_{sg}-\overline R_{L,g}\right)\right] {F}^{\rm BDH}_{L,\rm LO}(x,Q^2) \right .
\nonumber \\&& \left .
- a_s^2(Q^2)
\left[\frac{1}{3} \hat{B}^{(1)}_{L,s}L_A
+ \overline{B}^{(1)}_{L,s}(1) \right]
F^{\rm BDH}_2(x,Q^2)\right\}.\label{FL.8n}
\eea

This is our final expression for the longitudinal SF ${F}^{\rm BDH}_L(x,Q^2)$ within the NLO approximation
for low values of $x$.

\section{Results}\label{sec:results}
With the explicit form of the basic expressions  laid above, we can proceed
to extract  the longitudinal structure function
$F_L(x,Q^2)$ from data mediated by the BDH-parametrization of $F_2^{BDH}(x,Q^2)$.
In our calculations we employ   the standard representation for QCD couplings in the LO and NLO
(within the $\overline{MS}$-scheme) approximations
\bea
\begin{array}{lll}
a_s(Q^2) = \displaystyle\frac{1}{\beta_0 \ln\Bigl(Q^2/\Lambda^2\Bigr)} &\quad & {\rm  (LO),} \label{LO}\\
a_s(Q^2) = \displaystyle\frac{1}{\beta_0 \ln\Bigl(Q^2/\Lambda^2\Bigr)} -
 \displaystyle\frac{\beta_1 \ln \ln\left( Q^2/\Lambda^2\right)}{\beta_0 \left[\beta_0 \ln\Bigl(Q^2/\Lambda^2\Bigr)\right ]^2}
&\quad & {\rm(NLO),} \label{NLO}
\end{array}
\eea


The QCD   parameter $\Lambda$ has been
extracted from  the running coupling $\alpha_s$ normalized at  the $Z$-boson mass,
$\alpha_s(M_Z^2)$,  using the $b$- and $c$-quarks thresholds according to Ref.~\cite{Chetyrkin:1997sg}.
Applying this  procedure to ZEUS data, with $\alpha_s(M_Z^2) = 0.1166$ \cite{Chekanov:2001qu},
we obtain the following results for   $\Lambda$, cf. Ref.~\cite{Illarionov:2004nw}
\bea
\begin{array}{rllllll}
{\rm LO}: & \Lambda(f= 5)& = 80.80\ \mbox{MeV},&\Lambda(f= 4)& = 136.8 \ \mbox{MeV}, &\Lambda(f= 3) = 136.8\ \mbox{MeV},\label{LambdaLO}\\
{\rm NLO}:& \Lambda(f= 5)& = 195.7\ \mbox{MeV},&\Lambda(f= 4) &= 284.0\ \mbox{MeV}, &\Lambda(f= 3) = 347.2\ \mbox{MeV}.\label{LambdaMS}
\end{array}
\eea

\begin{figure}[t]
\includegraphics[height=0.25\textheight,width=.6\hsize]{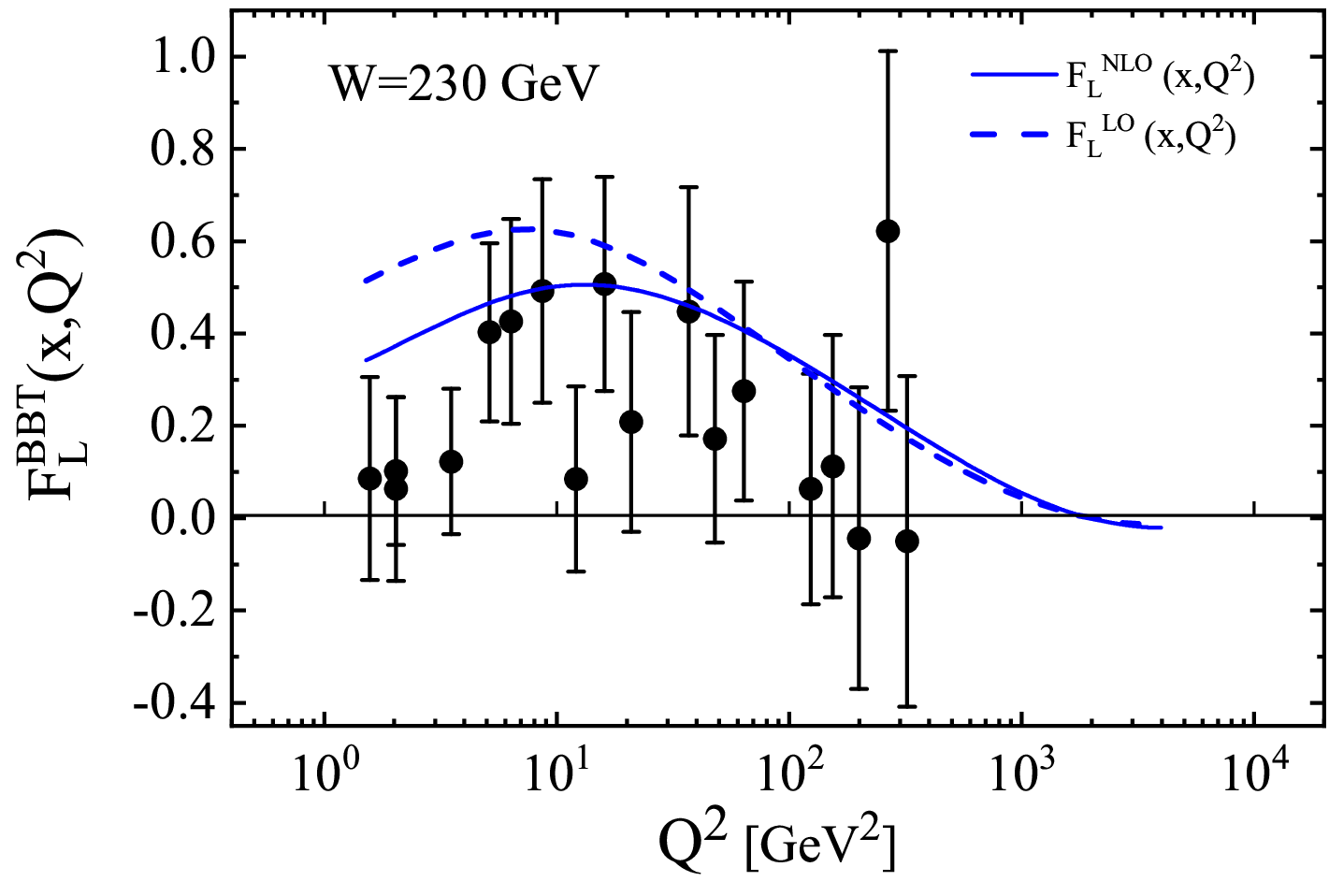}
\caption{(Color online)  The extracted longitudinal structure function
$F_L(x,Q^2)$ from the BDH-parametrization of $F_2(x,Q^2)$ at fixed value of the invariant mass $W=230$ GeV.
Dashed line -
calculations within the LO approximation, Eq.~(\ref{n9.01}), solid line - structure function
within the NLO approximation, Eq.~(\ref{FL.8n}).
 Experimental data are from the H1-Collaboration, Ref.~\protect\cite{Andreev:2013vha}. The Bjorken variable $x$
corresponding to the chosen kinematics  lies in the interval $(3\cdot 10^{-5}\,<\, x\, < 7\cdot 10^{-2})$.}
\label{fig:1}
\end{figure}

We have calculated the $Q^2$-dependence, at low $x$, of  the  longitudinal structure
function $F_L^{BDH}(x,Q^2)$  as described above, in the LO,
Eq.~(\ref{n9.01}), and NLO, Eq.~(\ref{FL.8n}),
approximations.
  Results of calculations  and comparison with data of the H1-Collaboration~\cite{Andreev:2013vha}
are presented in Fig.~\ref{fig:1}, where the dashed and solid lines correspond
to the extracted SF in the LO and NLO approximations, respectively.
Calculations have been performed at fixed value of the  invariant mass $W$,
$W=230$~GeV, allowing the Bjorken variable $x$ to vary in the interval $(3\cdot 10^{-5}\,<\, x\, < 7\cdot 10^{-2})$
when $Q^2$ varies in the interval $(1\, {\rm GeV^2}\,<\, Q^2\, < 3000\,{\rm GeV^2})$.
 Figure~\ref{fig:1} clearly demonstrates that the extraction procedure provides correct
behaviors of the extracted SF in both, LO and NLO approximations. At intermediate and high  $Q^2$ the
 extracted SF's are in a good agreement with experimental data. In this region the NLO corrections are rather small
 and can be neglected. A different situation occurs at low $Q^2 <\, 5\, {\rm GeV}^2$, where the LO $F_L(x,Q^2)$
substantially  exceeds experimental data. The NLO corrections here are negative and result in a better agreement
with data. However, at extremely low momenta, $Q^2 < 1.5\, {\rm GeV}^2$, the extracted SF within NLO
is still above the experimental data. It should also be mentioned that
our calculations are   consistent with other theoretical results, obtained, e.g.
in the framework of perturbation theory~\cite{Ball:2017otu,Abdolmaleki:2018jln} and/or
in the Ref. \cite{Kotikov:2004uf} in the so-called
$k_t$-factorization approach~\cite{Catani:1990xk}, both of
which incorporate  the BFKL
resummation ~\cite{Fadin:1975cb} at low $x$
(for a review of low-x  phenomenology see, e.g. cf.~Ref.~\cite{Andersson:2002cf}).
Recall that inclusion of the BFKL ressummation in study of PDF's lead to an improvement of
the description of data at small $x$ and nowadays appears as an integral part in a bulk of approaches.
The NNPDF~\cite{Ball:2017nwa} Collaboration and the
xFitter HERAPDF team~\cite{Aaron:2009aa,Alekhin:2014irh}, whose approaches are based on the DGLAP equations,
 recently included the BFKL resummation in to their analysis of the
combined H1$\&$ZEUS inclusive cross-section~\cite{Abramowicz:2015mha} achieving,
in such a way, a much  better description of data~\cite{Ball:2017otu,Abdolmaleki:2018jln}.
Analogous studies have been performed  in Refs.~\cite{Bonvini:2016wki,Ball:2017otu,Caola:2009iy,Bonvini:2019wxf}.
This is in some contrast with  results of  the standard PDF
sets~\cite{Jimenez-Delgado:2014twa,Dulat:2015mca,Ball:2017nwa,Alekhin:2014irh} without the BFKL resummations.

\begin{figure}[t]
\includegraphics[height=0.28\textheight,width=.6\hsize]{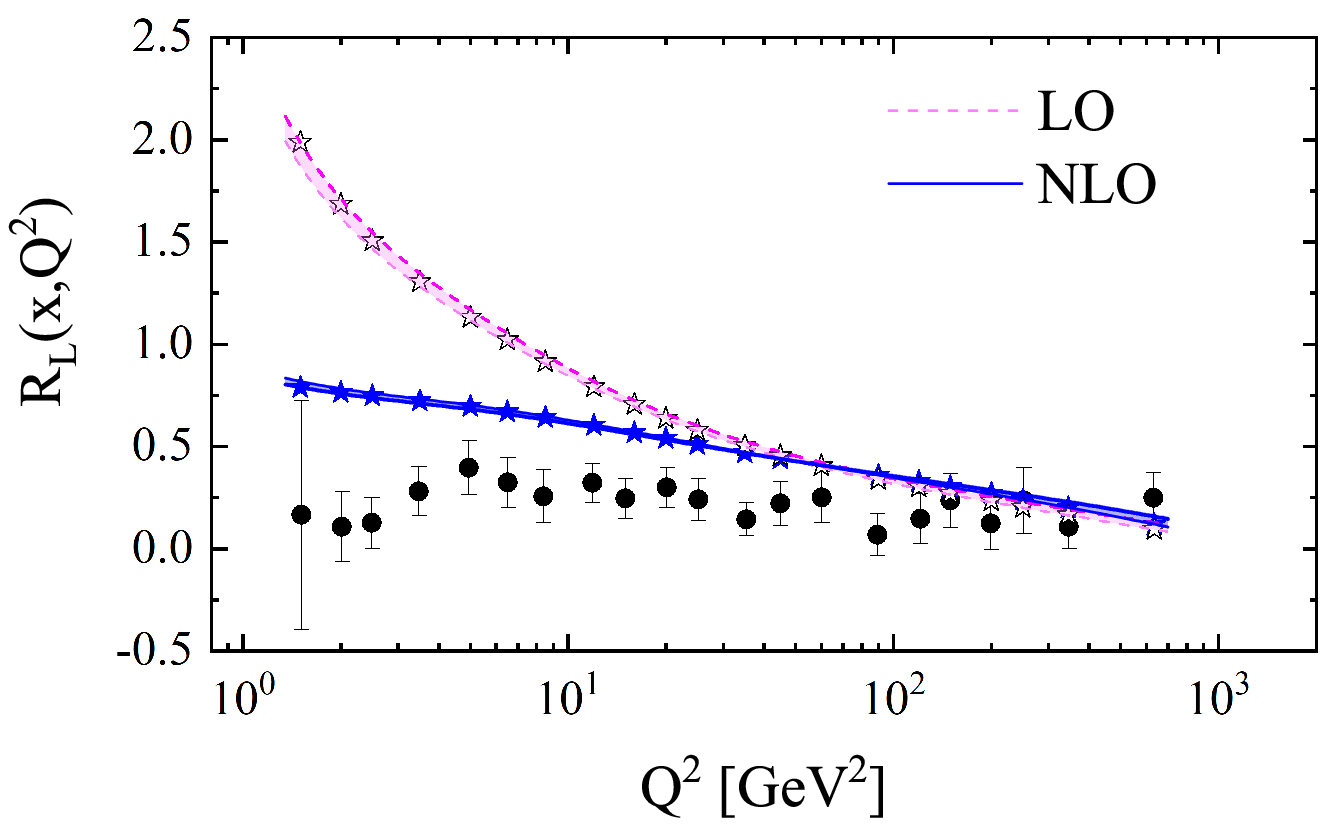}
\caption{(Color online)  The ratio of the longitudinal to transversal cross sections,
Eq.~(\ref{RL}), calculated with the extracted longitudinal SF within the leading  and
 next-to-leading order approximations. The open and full stars are results of LO and NLO
 calculations, respectively, within the exact kinematical
 conditions reported in Ref.~\cite{Andreev:2013vha}, i.e.  for each experimental
 point the variable $x$  is taken  from the corresponding  $(Q^2,x)$-bin. The shaded areas
are calculations
 with minimal and maximal values of $W$  from the Table~6 of Ref.~\cite{Andreev:2013vha},
 $W=232$~GeV and $W=184$~GeV for the upper and lower boundaries, respectively.}
\label{fig:2}
\end{figure}

 A particular interests present the ratio of the longitudinal to transversal cross sections, defined as
 \be
 R_L(x,Q^2) = \displaystyle\frac{F_L(x,Q^2)}{F_2(x,Q^2)-F_L(x,Q^2)}. \label{RL}
 \ee
Recently, the H1-Collaboration has reported the ratio $R_L(x,Q^2)$ measured in several kinematical
bins of averaged $Q^2$ and $x$, cf. Table~6 of Ref.~\cite{Andreev:2013vha}.
Within such kinematics, the invariant
mass $W$  changes  from $W\sim 230$~GeV to $W\sim 184$~GeV with increase of
$Q^2$ and $x$ in the selected bins. In Figure~\ref{fig:2}
we present  the ratio  (\ref{RL}), calculated with the extracted $F_L^{BDH}(x,Q^2)$  and
parametrized $F_2^{BDH}(x,Q^2)$, in comparison with the mentioned    H1-data.
The open and full stars are results of calculations within the LO and NLO approximations, where
$x$ and $Q^2$ correspond exactly
to  the experimental bins reported in Ref.~\cite{Andreev:2013vha}.
The shaded areas are calculations for two fixed, minimal and maximal,
 values of the invariant mass $W$
 within the chosen bins. From   Figure~\ref{fig:2} one can infer that
the NLO results essentially improve the agreement with data in comparison with the LO calculations.
As in the previous
case, the extracted longitudinal SF $F_L^{BDH}(x,Q^2)$ slightly overestimates the data at relatively low   $Q^2$.

\begin{figure}[t]
\includegraphics[height=0.35\textheight,width=.75\hsize]{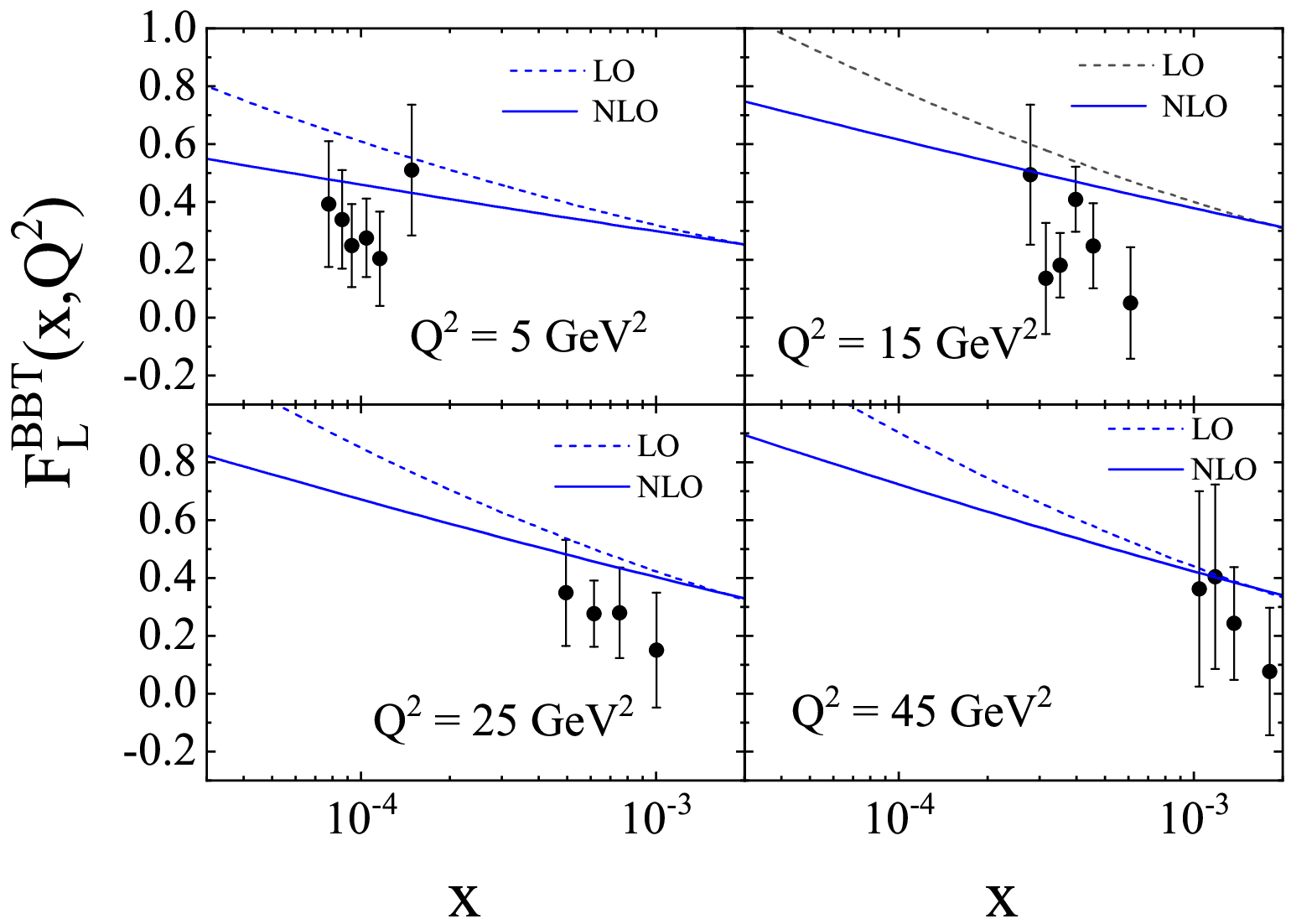}
\caption{(Color online)  The longitudinal structure function $F_L(x,Q^2)$ extracted
from the BDH-parametrization of $F_2(x,Q^2)$ at fixed $Q^2$ as a function
of the Bjorken variable $x$. The dashed lines represent results of calculations within the LO approximation,
the solid lines represent the SF obtained within the  NLO
approximation. Experimental  data are from the H1-Collaboration~\cite{Andreev:2013vha}.}
\label{fig:3}
\end{figure}

Now we proceed with an analysis of the x-evolution of the longitudinal SF at fixed $Q^2$.
As mentioned above, investigation of  $F_L(x,Q^2)$ as a function of $x$ is of interests
in connection with theoretical investigations of ultra-high energy processes with cosmic neutrinos and also
 in the context of the Froissart restrictions at $x\to 0$. We have calculated the $x$-dependence
of the longitudinal SF at several fixed values of $Q^2$ corresponding to H1-Collaboration data.
Results are presented in Figure~\ref{fig:3} where the $x$-evolution of $F_L(x,Q^2)$  is clearly exhibited.
It is seen that, for all values of the presented $Q^2$, the   extracted SF within the NLO
approximation is in a much better agreement with data. This persuades us that the obtained SF in the NLO approximation  can be pertinent in future analysis of the ultra-high energy neutrino data.

\begin{figure}[t]
\includegraphics[height=0.35\textheight,width=.7\hsize]{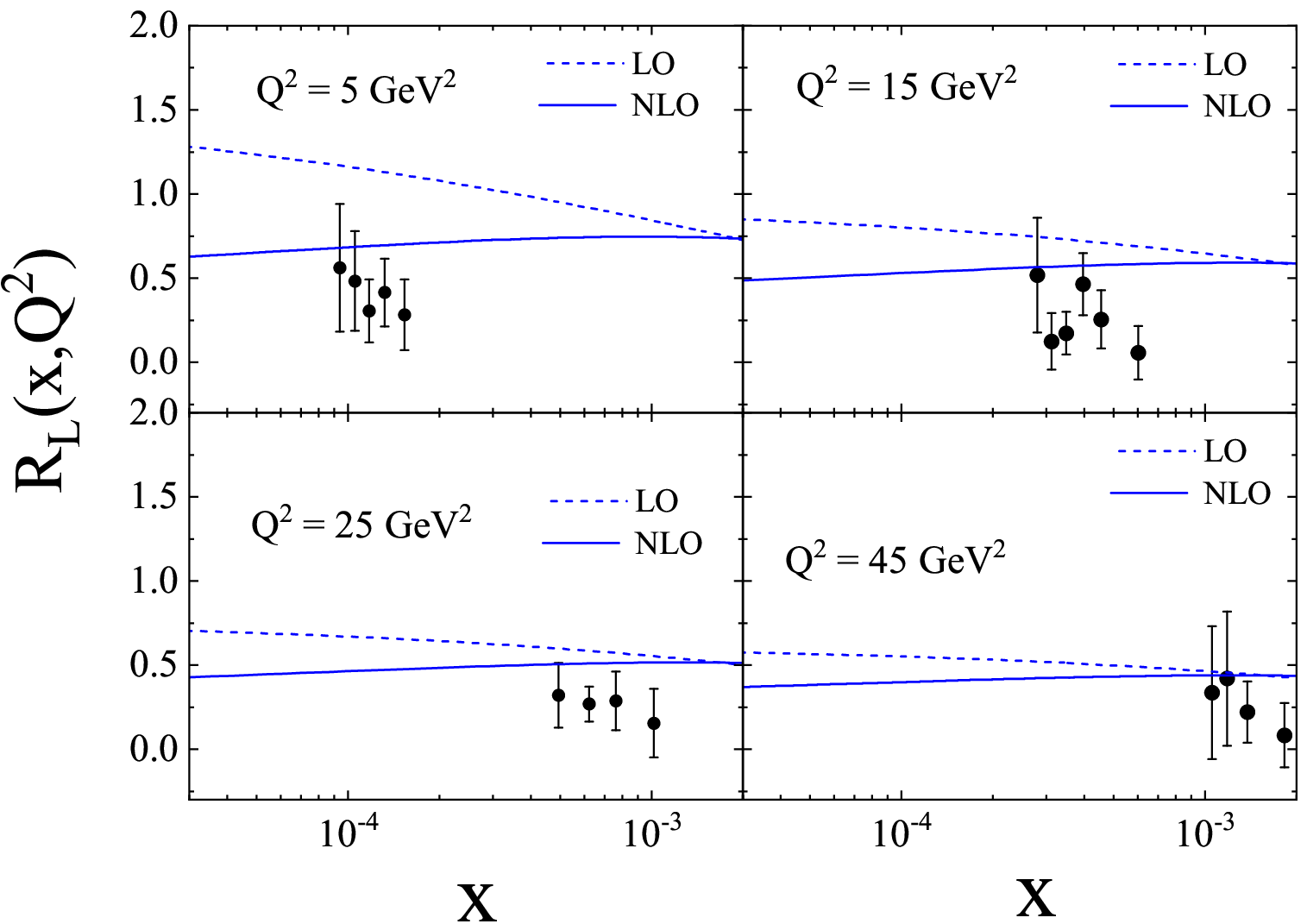}
\caption{(Color online)   }
\label{fig:4}
\end{figure}

In Figure~\ref{fig:4} we present the ratio~(\ref{RL}) calculated for the same kinematics as in Figure~\ref{fig:3}. As in previous calculations, the NLO results are in a better agreement with data. It is also seen
from Figure~\ref{fig:4}   that
the NLO ratio $R_L(x,Q^2)$ exhibits a tendency to be almost independent on $x$ in each bin of $Q^2$,
decreasing, however, with $Q^2$ increase, as it should be.
\noindent
We also mention that, as in the case of $Q^2$-dependence, our extracted  longitudinal SF
as a function of  $x$  is in a reasonable good agreement with
other theoretical predictions, see e.g. \cite{Ball:2017otu,Abdolmaleki:2018jln} for pQCD results and/or
Ref.~\cite{Kotikov:2002nh} for results obtained within the
 $k_t$-factorization approach.
Likewise  our results are in a good agreement with the previous
investigations reported in Ref.~\cite{Kotikov:1994jh},   were
a similar analysis has been performed within the framework of pQCD,
with  the experimental data for the transverse SF $F_2(x,Q^2)$
and the logarithmic derivative  $dF_2/d\ln(Q^2)$.

\section{Conclusions and outlook} \label{sec:con}
In this paper, we present a further development of the method of extraction of the
longitudinal DIS structure function $F_L(x,Q^2)$ suggested in Refs.~\cite{Kotikov:1994vb,Kotikov:1994jh,Kaptari:2018sxh}.
 The method relies on the Dokshitzer-Gribov-Lipatov-Altarelli-Parisi (DGLAP) equations and on the
Froissart-bounded parametrization of the DIS structure function $F_2(x,Q^2)$. We focus our attention
on the kinematical region of low Bjorken variable $(10^{-5}\,  \lesssim  \,x\, \, \lesssim 0.1)$ in a
 large interval of the momentum transfer  $(1\,{\rm GeV}^2 \, \lesssim  Q^2\,\lesssim \,3\cdot 10^3\,{\rm GeV}^2) $.
The extraction procedure has been elaborated for an analysis of the SF $F_L(x,Q^2)$ within the leading
and next-to-leading order approximations. To this end, we consider the transversal SF $F_2(x,Q^2)$ as known and use the
DGLAP equations to relate it to the longitudinal SF. Then, in space of Mellin momenta we find,
 up to $\alpha_s^2$ corrections, the corresponding Mellin transforms for the momenta corresponding
 to low and ultra-low values of $x$. The inverse Mellin transform
provides the sought  longitudinal SF in the usual $x-Q^2$ representation. The obtained explicit expression
for $F_L(x,Q^2)$ is entirely determined by the effective parameters of the
BDH partametrization~(\ref{parametrizationBDH}) and  is presented in Eq.~(\ref{FL.8n}).
Some comments are in order here. Observe that, the final expression~(\ref{FL.8n}) contains the
dominant logarithmic terms $\sim \ln(1/x)$ in both the numerator and denominator parts.
In principle, due to smallness of the running coupling
$\alpha_s(x,Q^2)$, the denominator part with $L_C$ can be rewritten in the numerator as an
alternate series leading to a behaviour  similar to one known within the pQCD, where
 the NLO corrections, in the considered kinematical region,
are negative and large, while NNLO contributions are positive and also large~(see, e.g. Ref.~\cite{Abt:2016vjh}).
A resummation of these contribution would allow to avoid such an alternate behaviour. In our case, this is achieved by
keeping the logarithmic term in the denominator without expanding it in to series relative to $\alpha_s(x,Q^2)$.
 With some extent, our representation of  the basic NLO corrections, Eq.~(\ref{FL.8n}), can be considered as an
effective  resummation of the  most  important, at low $x$,  logarithmic terms  in each order of the perturbation theory.
Another observation is that by keeping the logarithmic $L_C$ terms in the denominator
we also  manifestly demonstrate that the SF $F_L(x,Q^2)$ obeys the Froissart conditions.
As shown in Appendix~\ref{app:mel}, the logarithmic part in Eq.~(\ref{FL.8n}) is
a direct consequence  of the inverse Mellin transforms of terms $\sim 1/\omega$.

We have applied the developed method to extract the longitudinal SF within the kinematical conditions
corresponding to  that available at the HERA collider.
It has been found that, at relatively large $Q^2>10$~GeV$^2$
both, LO and NLO results reproduce fairly well the experimental data. At smaller $Q^2$ the LO approximation
fails to describe data, being systematically well above. Accounting for NLO corrections,
 which at low $x$ turn out to be negative, substantially improve the description of
  the SF and the ratio of the longitudinal to transversal cross sections. However, at
extremely low momentum transfer $Q^2 \lesssim1$~GeV$^2$, the extracted SF  still exceeds data.

We have performed an analysis of the $x$-evolution of the extracted SF.
It has been demonstrated that the $x$-dependence of $F_L(x,Q^2)$  also reproduces the behavior
of the experimental data at low $x$. Both, the extracted SF and the ratio $R_L(x,Q^2)$ as functions of $x$,
are in a fairly good agreement with data, herewith the NLO results are in a much better agreement not
only with data, but also with other
 existing theoretical investigations  based on perturbative QCD,   improved by the BFKL resummation
  (see Ref.~\cite{Ball:2017otu,Abdolmaleki:2018jln} and references therein quoted), as well as with the results~\cite{Kotikov:2004uf}
obtained in the framework of the  $k_t$-factorization method~\cite{Catani:1990xk},
also based on the BFKL approach~\cite{Fadin:1975cb}. Calculations of the longitudinal SF
based on  the traditional pQCD without such improvements  turn out to be rather unstable,
due to the fact that the subsequent perturbative corrections can be even larger than the
previous ones~\cite{Abt:2016vjh,Gauld:2016kpd}. Incorporation of corrections inspired by
 the BFKL resummation lead to a substantially more  stable results for $F_L(x,Q^2)$~\cite{Gauld:2016kpd} (see
also similar investigations in Refs.~\cite{Bonvini:2016wki,Ball:2017otu,Caola:2009iy,Bonvini:2019wxf}),
aimed to achieve a good description of  the combined H1$\&$ZEUS  inclusive cross-sections~\cite{Abramowicz:2015mha}.

Apart from the study of the Froissart
boundary restrictions, the knowledge of the $F_L(x,Q^2)$ at low $x$ is of a great interest in
connection with the theoretical treatments of the  ultra-high energy processes with  cosmic neutrinos.
As   already mentioned  in Introduction,
the NLO approximation for  $F_L(x,Q^2)$,
i.e. calculations up to $\alpha_s^2$-corrections,
corresponds to  next-next-to-leading order (NNLO) for $F_2(x,Q^2)$  which,  in the LO, is $\propto \alpha_s^0$.
 Consequently, with the NLO results (\ref{FL.8n}),
 in our approach it becomes possible  to perform NLO and NNLO analyses of the ultra-high energy ($\sqrt{s}\sim 1$~TeV)
 neutrino cross-sections similar to  NLO \cite{CooperSarkar:2007cv} and  NNLO ~\cite{Bertone:2018dse}
 investigations based on pQCD.   Such   calculations are of a great importance in view of  expected
 reliable cross sections  from existing and  forthcoming data at the IceCube~\cite{Bustamante:2017xuy}  and
 from the substantially improved  IceCube-Gen2~\cite{Aartsen:2014njl} Collaborations.
Therefore,  a direct  comparison of the theoretical predictions with experimental data becomes feasible.

  In the kinematical region where the gluon contributions are sizable
the  (large) corrections within the traditional pQCD  can be strongly  reduced
by a proper  change of the factorization and renormalization scales~\cite{Kotikov:1993yw}.
 An analysis of the precise H1$\&$ZEUS combined  data~\cite{Aaron:2009aa} obtained
  within the kinematics near the
 limit of applicability of pQCD has shown~\cite{Kotikov:2012sm} that
 an employ of effective scales with large parameters provides much smaller
 high-order perturbative corrections. In such a case  the  strong couplings decrease as well and, as a rule,
calculations with effective scales lead to a better agreement with data
(cf.  investigations of the NLO corrections in context with high-energy asymptotics
of virtual photon-photon collision~\cite{Brodsky:1998kn}  and studies of  $F_L(x,Q^2)$ in the framework of
the $k_t$-fragmentation approach~\cite{Kotikov:2004uf}). This encourages us to continue our
low-$x$ analysis of the SF's  by implementing special change of the factorization
and renormalization scales~\cite{Kotikov:1993yw}. This is the subject of our further investigations and
results will be presented elsewhere.


  Furthermore,    we   plan to improve our approach  by accommodating the method
  to extract, in the  LO and NLO approximations, also the gluon densities.
  We shall note that, the gluon distribution is by far less known, experimentally and
    theoretically. Even the shape of the gluon density  are often taken quite different in different
  PDF sets~\cite{Dulat:2015mca,deOliveira:2017lop}, although considered within the same prerequisites of pQCD.
  However, the range of variation of gluon density strongly decreases when BFKL resummation is included
in the analyses at low $x$  (see the most recent publication~\cite{Bertone:2018dse} and discussion therein).

 An extraction  of the gluon distributions from experimental data,
 performed within an approach similar to   the one suggested in the present paper,
 can provide valuable additional information on the problem. Such an analysis can be accomplished
by  employing  the charm, $F_2^{cc}(x,Q^2)$, and beauty, $F_2^{bb}(x,Q^2)$,
components of the  SF $F_2(x,Q^2)$, which are directly related
to the gluon density in the  photo-gluon fusion reactions   (see Ref. \cite{Illarionov:2008be} and discussion therein).
The extracted SF's can be compared with the recently  obtained    combined data
from the H1$\&$ZEUS-Collaboration~\cite{H1:2018flt} for the  $F_2^{cc}(x,Q^2)$ and $F_2^{bb}(x,Q^2)$
  and with the theoretical predictions~\cite{Bertone:2018dse} based on pQCD with
 BFKL corrections included, and
also with the results~\cite{Kotikov:2001ct} obtained in the framework of the $k_t$-fragmentation.

Furthermore, the charmed parts of transverse  $F_2^{cc}(x,Q^2)$ and longitudinal, $F_L^{cc}(x,Q^2)$, structure functions
being calculated   within our approach, can be used to predict the
charmed part  of the neutrino-nucleon cross-sections at ultra-high energy
and to compare    with other  calculations~\cite{Bertone:2018dse}
based on pQCD with BFKL corrections. Yet, the BDH-gluon density itself can serve as a
useful tool  for  estimations of  the cross sections with cosmic rays, cf. Ref.~\cite{Reno:2019jtr}
(for  a most recent review on the subject,
 see Ref.~\cite{Anchordoqui:2018qom} and references therein quoted).
  Recall that,  the gluon density in the  BDH-like  form already contains
 information about violation of the standard DGLAP
 evolution (cf. discussions in Sec. II) and indicates on   possible presence of shadowing effects,
 which are among the  basic subjects of physical programs of operating (e.g. NICA in Dubna) and planned
facilities (EIC@China, ELIC@JLAB, ENC@GSI \ldots) aimed on study of properties of nuclear matter at high energies~\cite{Accardi:2012qut}. Our investigations in this direction are in progress.

In summary, we present a theoretical method to extract, from the experimental data,
 the longitudinal DIS structure function $F_L(x,Q^2)$ at low $x$ within the leading and next-to-leading
  order approximations. Explicit, analytical expressions for the structure function in both, LO and NLO, approximations
 are obtained in terms of the effective parameters of the Froissart-bound parametrization of $F_2(x,Q^2)$
  and results of numerical calculations as well as comparisons with available experimental are presented.

\acknowledgments
Support by the National Natural Science Foundation of China (Grants
No. 11575254) and the National Key Research and Development Program of China (No. 2016YFE0130800) is acknowledged.
LPK and AVK highly appreciate the warm hospitality
at the Institute of Modern Physics
and thank the CAS President's International Fellowship Initiative
(Grant  No.~2018VMA0029 and
No.~2017VMA0040) for support.

\appendix

\section{D\lowercase{etails of evaluation of some relevant integrals} }
\label{App:B}

Herebelow we present the evaluation of the integrals $P_k(\omega,\nu)$ $(k=0,1,2)$ appearing
 in Eq. (\ref{IntMain}). It can be rewritten in a more general form

 \be
 \hat{P}_k(\omega,\nu) = \int^{1}_0 dx x^{\omega-1} (1-x)^{\nu} \left(\ln \frac{1}{x}\right)^k
=  \left(-\frac{d}{d\omega} \right)^k \, \int^{1}_0 dx x^{\omega-1} (1-x)^{\nu}  \, ,
 ~~ (k=0,1,2) .
\label{IntBa}
\ee

\subsection{$k=0$}
\be
 \hat{P}_0(\omega,\nu) = \int^{1}_0 dx x^{\omega-1} (1-x)^{\nu}  = \frac{\Gamma(\omega)\Gamma(\nu +1)}{\Gamma(\omega+\nu +1)}
 = \frac{1}{\omega} \, \frac{\Gamma(\omega+1)\Gamma(\nu +1)}{\Gamma(\omega+\nu +1)} \, .
\label{Int0}
\ee

The last results in the r.h.s. can be represented as (cf. also~\cite{Fleischer:1998nb})
\be
\frac{\Gamma(\omega+1)\Gamma(\nu +1)}{\Gamma(\omega+\nu +1)} = \exp \Bigl[-\sum_{i=1}^{\infty} S_i(\nu) \omega^i\Bigr] \, ,
\label{Sum}
\ee
where the nested sums  $S_i(\nu)$ are defined by Eqs.~(\ref{Sm}) and (\ref{-Sm}).

Expanding r.h.s. of (\ref{Int0}) in $\omega$ series, we have
\be
 \hat{P}_0(\omega,\nu)  = \frac{1}{\omega}  -S_1(\nu) \, .
\label{Int0.1}
\ee

\subsection{$k=1,2$}
\label{app:c}
For the next basic integral $\hat{P}_1(\omega)$ we have
\be
 \hat{P}_1(\omega,\nu) = \left(-\frac{d}{d\omega} \right)\, \hat{P}_0(\omega,\nu) = \left(-\frac{d}{d\omega} \right)\,
  \frac{1}{\omega} \, \frac{\Gamma(\omega+1)\Gamma(\nu +1)}{\Gamma(\omega+\nu +1)} \, .
\label{Int1}
\ee

Expanding r.h.s. of (\ref{Int1}) in  series w.r.t. $\omega$, we have
\be
 \hat{P}_1(\omega,\nu)  = \frac{1}{\omega^2}  -Z_2(\nu) \, ,
\label{Int1.1}
\ee
where (see, for example, \cite{Cvetic:2011ym})
\be
Z_1(\nu)=S_1(\nu),~~Z_2(\nu)=\frac{1}{2} S_1^2(\nu) - \frac{1}{2} S_2(\nu), ~~
Z_3(\nu)=\frac{1}{6} S_1^2(\nu) - \frac{1}{2} S_1(\nu) S_2(\nu) + S_3(\nu),
\label{Zi}
\ee
where $S_i(\nu)$, for integer $\nu$,  are the known
harmonic numbers $S_i(\nu)=\sum_{k=1}^{\nu} 1/k^i$. For arbitrary arguments $\nu$, these coefficients
  are related to the Euler $\Psi(1+\nu)$-function and
its derivatives $\Psi^{(m)}(1+\nu)=d/(d\nu)\Psi(1+\nu)$ as
\be
S_1(\nu)= \Psi(1+\nu) +\gamma_{\rm E},~~S_2(\nu)= \zeta_2 - \Psi^{(1)}(1+\nu),~~
S_2(\nu)=\frac{1}{2} \, \Bigl(\Psi^{(2)}(1+\nu) - \zeta_3 \Bigr) \, ,
\label{Si}
\ee
where $\gamma_{\rm E}$ is Euler constant and $\zeta_i$ are Euler $\zeta$-functions.

Analogous calculations  for $\hat{P}_2(\omega)$ provide
\be
 \hat{P}_2(\omega,\nu) = \left(-\frac{d}{d\omega} \right)^2 \, \hat{P}_0(\omega,\nu) = \left(-\frac{d}{d\omega} \right)^2\,
  \frac{1}{\omega} \, \frac{\Gamma(\omega+1)\Gamma(\nu +1)}{\Gamma(\omega+\nu +1)} \, ,
\label{Int2}
\ee
which, being expanded in to series about $\omega$, results in
\be
 \hat{P}_2(\omega,\nu)  = 2 \, \left(\frac{1}{\omega^3}  -Z_3(\nu) \right)\, .
\label{Int2.1}
\ee

Equations (\ref{Int0.1}), (\ref{Int1.1}) and (\ref{Int2.1}) allow to write the considered integral
(\ref{M2BDHsing}) as
\be \int^{1}_0 dx x^{\omega-1} (1-x)^{\nu} L^k(x) = P_k(\omega,\nu) + O(\omega) \, ,
\ee
where
\bea
&&P_0(\omega,\nu)=\frac{1}{\omega} - Z_1(\nu) \, ,~~
P_1(\omega,\nu,L_1)= \frac{1}{\omega^2} - Z_2(\nu) + L_1 P_0(\omega,\nu) \, ,  \nonumber \\
&&P_2(\omega,\nu,L_1)= 2 \left(\frac{1}{\omega^3} - Z_3(\nu)\right)  + 2L_1 P_1(\omega,\nu) +
L_1^2 P_0(\omega,\nu) \, ,
\label{Pi}
\eea
In Eq.~(\ref{Pi}) the finite part of the integral is encoded in functions $Z(\nu)$, Eqs.~(\ref{Zi}), while the
 singular one is given by Eqs.~(\ref{PiSing}).

\subsection{$k=3$}


Consider now the integral
\be \int^{1}_0 dx x^{\omega-1} (1-x)^{\nu} \left(\ln \frac{1}{x}\right)^3 = P_3(\omega,\nu) + O(\omega) \, ,
\label{Int}
\ee
which is the third derivative of $\hat{P}_3(\omega)$ with respect to $\omega$
\be
 \hat{P}_3(\omega,\nu) = \left(-\frac{d}{d\omega} \right)^3 \, \hat{P}_0(\omega,\nu) = \left(-\frac{d}{d\omega} \right)^2\,
  \frac{1}{\omega} \, \frac{\Gamma(\omega+1)\Gamma(\nu +1)}{\Gamma(\omega+\nu +1)} \, .
\label{Int1C}
\ee

Eventually, at $\omega\to 0$ we obtain
\be
 \hat{P}_3(\omega,\nu)  = 6 \, \left(\frac{1}{\omega^4} + O(\omega^0)
 \right)\, .
\label{Int2C}
\ee

\section{I\lowercase{nverse} M\lowercase{ellin transforms at low $x$}}
\label{app:mel}
In this  section we present some details of calculations of the inverse Mellin transform
of the longitudinal momentum $M_L(n,Q^2)$, Eq.~(\ref{FL.6n}). Observe, that
$M_L(n,Q^2)$ is expressed via $M_L^{LO}(n,Q^2)$ and $M_2^{BDH}(n,Q^2)$. Hence, it is sufficient to
determine the inverse Mellin transforms of $M_2^{BDH}(n,Q^2)$ augmented with some coefficients depending
on $\omega$ to find the desired longitudinal SF $F_L(x,Q^2)$.
To facilitate the calculations,
consider the following  auxiliary   integral

\be
I(\omega,Q^2)=\int^{1}_0 dx x^{\omega-1} \left(\ln \frac{1}{x}\right) F_{2}^{\rm BDH}(x,Q^2).
\label{aux}
\ee
It is obvious that this integral is proportional to the Mellin transform $M_2(\omega,Q^2)$ with some coefficients of
proportionality as functions of  $\omega$,
\be
I(\omega,Q^2)\propto \left( \frac{K_{-1}}{\omega}+K_0+K_1 \omega + K_2 \omega^2 +\cdots \right)
M_2^{BDH}(\omega,Q^2).
\label{aux1}
\ee
If so, we can avoid direct calculations of the inverse  Mellin transform of $M_2^{BDH}(\omega,Q^2)$.
Instead, for any constants $\hat F_1$ and $\hat F_2$ and vanishing $\omega$,  we can use the obvious relation
\be
\left(\frac{\hat F_1}{\omega}+\hat F_2\right )M_2^{BDH}(\omega,Q^2)\overset{{\rm Inverse \, Mellin}}{\xrightarrow{\hspace*{2cm}}}
\left(\frac{\hat F_1}{3}L_A +\hat F_2\right) F_2^{BDH}(x,Q^2),
\ee
where $L_A= \ln \frac{1}{x} + L_1 + \frac{A_1}{2A_2}$, cf. Eqs.~(\ref{LB}) and~(\ref{dependencies}).
The integral $I(\omega,Q^2)$ in (\ref{aux}) can be calculated directly
by  sing Eqs. (\ref{Int0.1}), (\ref{Int1.1}), (\ref{Int2.1}) and (\ref{Int2C}).
Up to  ${\cal O}(\omega^0)$, we have
\be
\int^{1}_0 dx x^{\omega-1} \left(\ln \frac{1}{x}\right) F_{2}^{\rm BDH}(x,Q^2)
= D \Biggl[ A_0 \frac{1}{\omega^2} + A_1 \left(\frac{2}{\omega^3}+ \frac{L_1}{\omega^2}\right)
+ A_2 \left(\frac{6}{\omega^4}+\frac{4L_1}{\omega^3}+ \frac{L_1^2}{\omega^2}\right)\Biggr] .
\label{Int3C}
\ee
Expression (\ref{Int3C}) must be compared with Eq.~(\ref{aux1}), which, after
  insertion of (\ref{M2BDHsing})-(\ref{PiSing})
reads as
\bea
&&\left(\frac{K_{-1}}{\omega} + K_0 + K_1\omega + K_2\omega^2\right)  M_{2}^{\rm BDH}(\omega,Q^2)
\label{Int4C} \\
&&=\left(\frac{K_{-1}}{\omega} + K_0+ K_1\omega + K_2\omega^2\right) D \Biggl[ A_0 \frac{1}{\omega} + A_1 \left(\frac{1}{\omega^2}+ \frac{L_1}{\omega}\right)
+ A_2 \left(\frac{2}{\omega^3}+\frac{2L_1}{\omega^2}+ \frac{L_1^2}{\omega}\right)\Biggr]\, .  \nonumber
\eea

Now, equating  in (\ref{Int3C}) and (\ref{Int4C}) the corresponding
coefficients  in front of   $\omega^{-k}$  we obtain
\bea
&&K_{-1}=3,~~K_0=-L_1-\frac{A_1}{2A_2},~~K_1=-\frac{A_0}{A_2}+\frac{A_1^2}{4A_2^2},~~ \nonumber \\
&&K_2=\frac{1}{2}L_1^3+\frac{3A_1}{4A_2}L_1^2 +\frac{3A_0}{2A_2}L_1 + \frac{3A_0A_1}{4A_2^2}-\frac{A_1^3}{8A_2^3}.
\label{Int5C}
\eea



\end{document}